\documentclass{emulateapj}

\usepackage{amsmath}
\usepackage{graphicx}
\usepackage{subfigure}
\usepackage{multirow}

\begin{document}

\title{Stellar Metallicity Gradients in SDSS galaxies}

\author{
Benjamin Roig\altaffilmark{1},
Michael R. Blanton\altaffilmark{1},
Renbin Yan\altaffilmark{2}
}

\altaffiltext{1}{
  Center for Cosmology and Particle Physics, Department of Physics, New
  York University, 4 Washington Place, New
  York, NY 10003}
\altaffiltext{2}{
  University of Kentucky}

\begin{abstract}
We infer stellar metallicity and abundance ratio gradients for a
sample of red galaxies in the Sloan Digital Sky Survey (SDSS) Main
galaxy sample. Because this sample does not have multiple spectra at
various radii in a single galaxy, we measure these gradients
statistically. We separate galaxies into stellar mass bins, stack
their spectra in redshift bins, and calculate the measured absorption
line indices in projected annuli by differencing spectra in
neighboring redshift bins.  After determining the line indices, we use
stellar population modeling from the EZ\_Ages software to calculate
ages, metallicities, and abundance ratios within each annulus.  Our
data covers the central regions of these galaxies, out to slightly
higher than $1 R_{e}$. We find detectable gradients in metallicity and
relatively shallow gradients in abundance ratios, similar to results
found for direct measurements of individual galaxies.  The gradients
are only weakly dependent on stellar mass, and this dependence is
well-correlated with the change of $R_e$ with mass. Based on this
data, we report mean equivalent widths, metallicities, and abundance
ratios as a function of mass and velocity dispersion for SDSS
early-type galaxies, for fixed apertures of 2.5 kpc and of 0.5 $R_e$.
\end{abstract}

\keywords{}

\section{Introduction}

The study of stellar population and metallicity gradients has been an
ongoing topic because it places relatively strong constraints on the
evolutionary history of galaxies. This has been a much-studied field
for the past 50-60 years, with a wealth of information available that 
allows the interpretation of galaxy spectra in terms of stellar 
populations (\citealt{faber85a} for example). 
Early studies that correlated observable metal absorption lines with stellar
population properties only could examine small numbers of galaxies
(e.g. \citealt{couture88a, munn92a, gonzalez95a}), but led to the acceptance
that galaxies typically had lower metallicities at larger radii. Later
studies have increased the sample size, measured additional metal 
absorption indices, and added detail to that basic picture \citep{carollo93a,
koleva11a}.

Theoretical studies have also, of course, addressed the question of 
what causes this common trend in elliptical galaxies 
(\citealt{martinelli98a, ogando06a} are several of many). Mergers, 
star formation histories, gas flow, and other mechanisms help determine 
the metallicity gradients. Mergers can
drastically change the distribution of stars in the final galaxies,
affecting the observed metallicity gradients dramatically as well
\citep{dimatteo09a}. Conversely, a lack of mergers can lead to the
evolution of a metallicity gradient after the infall of cooling gas
\citep{pipino10a}.  Predicting the observed stellar metallicity
gradients correctly requires a model of the star formation and
formation history (mergers or otherwise) of galaxies. This suggests
that a large-scale study of metallicity gradients in elliptical
galaxies may help our understanding of the average path an elliptical
galaxy takes in formation --- how many mergers, mass ratios of the
mergers, when star formation bursts occur, and how long these
formation bursts last, among other properties.

Equally, observing metallicity gradients in a large sample of galaxies
has been a challenge, as it requires spectra of multiple regions of
the same galaxy. Usually, then, observational studies that attempt to
constrain theoretical models are restricted to a relatively small
number of nearby galaxies \citep{greene13a, spolaor10a, kuntschner10a,
  rawle10a, pastorello14a}, which makes them sensitive to the specific
choice of galaxies to observe. Newer projects (including 
\cite{gonzalezdelgado14a} with the CALIFA survey and \cite{bundy15a} 
with MaNGA) study these gradients using integral field spectroscopy (IFS)
to attempt to improve our understanding. Here, we study a 
large number of galaxies observed by the Sloan Digital Sky Survey (SDSS) 
and instead of finding metallicity gradients
in individual galaxies, we average galaxies in redshift bins and
calculate statistical gradients between annuli found by subtracting
galaxy fluxes at different redshifts from each other.  This process
yields the population-averaged metallicity and abundance profiles of
early type galaxies.

We begin by selecting a sample of non-starforming galaxies in SDSS, 
as detailed in Section \ref{sec:sampleselect}. We measure 
the standard Lick indices used to calculate age and metallicity, and make 
use of EZ\_AGES \citep{graves08a} to obtain those parameters. With our sample 
complete, we compare our calculated gradients to several other studies in 
Section \ref{sec:obs} to verify the accuracy of our approach, focusing on 
studies that also are able to find gradients in the inner $1 R_{e}$ of 
galaxies rather than ones
that examine the full outer regions as well. We then examine 
several papers that discuss theoretical models for the formation of these 
galaxies to see which our observations support and what, if any, conclusions 
we can draw about likely evolutionary histories of our galaxies in Section 
\ref{sec:theory}. 

\section{Sample Selection}

\label{sec:sampleselect}

Our sample is composed of a subset of the NYU Value-Added Galaxy
Catalog (NYU-VAGC; \citealt{blanton05a}), based on the SDSS Data
Release 7 (DR7; \citealt{abazajian09a}). We use 686,356 Main sample
galaxies (\citealt{strauss02a}) with observed spectra and several
extracted values from the observations (redshift, a half-light radius,
magnitudes in several bands, and an estimate of the stellar mass). The
stellar mass we use is estimated from the K-corrected mass-to-light 
ratios in the $ugriz$ and JHK bands (see \citealt{blanton07a} for 
details).

We wish to select a uniform sample of non-starforming galaxies so that
we can have confidence that the properties remain relatively the same
across the entire redshift range of our study and avoid any emission
line contamination of the features we seek to measure. To do this, we
perform a few simple cuts involving [OII] and H$\alpha$ equivalent
widths (EWs) as previously done in \cite{yan06a} that yield cuts in
spectroscopic properties.  First, objects that have [OII] emission but
no H$\alpha$ are kept in the sample. Second, any object without either
[OII] or H$\alpha$ emission is kept. Finally, for objects with both
[OII] and H$\alpha$, we keep ones with a high ratio of [OII]/$H\alpha$
and reject those with a low ratio. We define this ratio as
\cite{yan06a} does $({\rm EW}([OII]) > 5 {\rm EW}(H\alpha) - 7)$ and
we keep all objects that pass this ratio of those with both lines
detected. These cuts may potentially exclude some red, non-starforming
galaxies as the cost of being sure the number of blue, starforming
galaxies remaining in the sample is very low (approximately 3\% per
\citealt{yan06a}). We do not, however, make any morphology cuts; 
this suggests about 40\% of the sample will be pure elliptical galaxies, with
the rest a mix of S0 or Sa morphologies \citep{blanton09a}. The dependence 
of gradients on a mix of E, S0, and Sa morphologies has been found to have 
limited effect (see \citealt{gonzalezdelgado14a}, Fig. 5), so this may not 
dramatically impact our results, but should be remembered as a caveat.

This leaves us with 266,195 non-starforming galaxies to work with. The
next set of cuts is to ensure that the data remaining has correctly
measured results for all the important parameters that we will
need. We remove all objects with masses less than $10^{7}$ and all
objects that do not have measured EWs for the metal indices (Mg b,
Fe5270, Fe5335, Ca4227, and C4668) that we will be using, as well as
H$\beta$, which tends to happen only in the rare case of a badly fit
absorption line. This is a relatively minor adjustment and only
reduces our sample by approximately 4\%.

Finally, metallicities are dependent on the stellar masses of the
galaxies, and so we break up the data into mass bins, each
individually volume limited, the details of which are given in Table
\ref{table:massbins}. We have also alternatively divided the galaxies
into velocity dispersion bins, with the name number of bins and volume
limiting cuts in redshift and magnitude.

As a note, the mass bins are large enough to still leave around 
a factor 
of two difference in the brightest versus dimmest galaxies in each bin; 
however, in tests to see if this luminosity range overly weighted the
brightest galaxies in each bin in measuring line fluxes we found less
than a 5\% steepening of the indicator gradients when normalizing 
luminosities versus leaving them unchanged. For the results presented in
this paper, we do not normalize each galaxy in a mass bin to the same
luminosity, but rather leave the luminosities as they are for each
individual galaxy.

\begin{table}[t!]
  \centering
  \caption{Volume Limited Mass Binning of Galaxies in this Paper}
  \begin{tabular}{ c | c | c | c | c | c}
$\log{(M)}$ Range & z Range & Max V Mag & Objects \\
\hline
$10.0 < \log{M} < 10.3$ & $0.02 < z < 0.09$ & -19.35 & 23324 \\
$10.3 < \log{M} < 10.7$ & $0.02 < z < 0.12$ & -20.05 & 57381 \\
$10.7 < \log{M} < 11.0$ & $0.03 < z < 0.18$ & -21.05 & 50880 \\
$11.0 < \log{M} < 11.5$ & $0.07 < z < 0.24$ & -21.80 & 22029 \\
\end{tabular}  
\tablecomments{4 stellar mass bins are chosen for the galaxes in this paper 
to reduce the effects of the mass dependence of metallicity. Each bin has a 
redshift range 
and magnitude limit selected to maximize the number of objects in the sample 
while ensuring that the sample remains volume limited.}
  \label{table:massbins}
\end{table}

\begin{figure}[t!]
  \centering
  \includegraphics[scale=0.4]{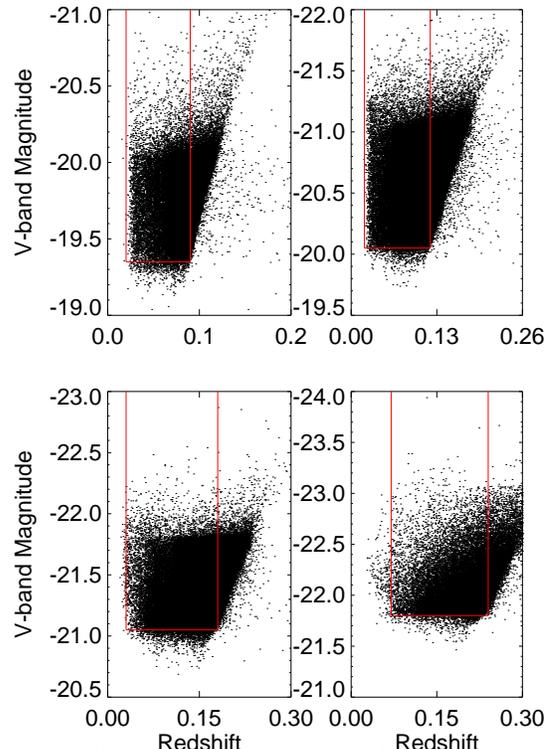}
  \caption{The cuts made to create a volume-limited sample for each stellar mass bin.}
  \label{fig:vollimit}
\end{figure}

Figure \ref{fig:vollimit} shows the resulting samples with the cuts in red 
lines.

In preparation for Lick index measurement, we must make two
adjustments to our spectra. The first is a resolution correction ---
as \cite{zhu10a} notes when performing the same corrections, SDSS
spectra are not at the same resolution at which the originally defined
Lick indices were observed. \cite{schiavon07a} contains several tables
(Tables 43--46, depending on galaxy age) in which corrections for a
given velocity dispersion to the Lick IDS resolution are defined, but
to establish as common a baseline as we can between mass bins, we
smooth all galaxies in our sample to the same velocity dispersion of
$325$ km s$^{-1}$. Any errors in the resolution correction from
\citet{schiavon07a} will thus at least be the same for every galaxy,
hopefully ensuring that any gradient trends we report will be
independent of these errors. We use the table of corrections for a 7.9
Gyr population, noting that while some of our galaxies will be older
and some younger, the differences are only a few percent for
corrections of a different stellar population age. The second
correction, also defined by \cite{schiavon07a}, is a fluxing
correction to bring the SDSS spectra in line with the Lick IDS
system. It too is empirical, with potentially large error, but we
apply this correction after each Lick index is measured to fully bring
our results in line with the Lick IDS system, noting unfortunately
that errors from these corrections cannot be included in our error
budget as well.

Each object that is kept in the sample then has its Lick indices for
Mg b, Fe 5335, Fe 5270, Ca 4227, C 4668, and H$\beta$ measured. In
addition to the Lick index, we also measure the absorption line flux
and continuum values directly over the same wavelength range. We also
measure the [OIII] line flux to correct for H$\beta$ line infill due
to emission (see \citealt{zhu10a, trager00a, mehlert00a} for some
discussion on the process and difficulties behind this correction). We
use the standard correction factor of $\Delta$EW H$\beta = 0.6 {\rm
  EW}([OIII])$ that these prior works settle on with the understanding
that it is empirical and has relatively large scatter, instead of
attempting a correction where we would directly adjust the flux
(instead of the equivalent width) of the H$\beta$ absorption based on
the measured flux of the [OIII] emission. In this [OIII] line fitting,
we ignore any lines that are weak detections (defined as
indistinguishable from zero line flux or EW at the one sigma
level), and simply do not correct H$\beta$ for those objects. 
We do have data for other commonly used indicators such as D4000
or CN1 and CN2, but the population modeling code we use
does not use these as inputs to determine metallicity or age, so we
do not present them here.

We convert several other of the above measured Lick indices to
commonly used combinations in calculations of metallicity as
well. First we compute $\mathrm{[MgFe]'=\sqrt{Mg b \times (0.72 Fe5270
    + 0.28 Fe5335)}}$, and secondly we compute $\mathrm{\langle
  Fe\rangle = (Fe5270 + Fe5335) / 2}$. Both these combinations are
chosen because they more directly correlate with metallicity than any
single Lick index \citep{thomas03a}. This gives us 5 metallicity
indicators ($\mathrm{[MgFe]'}$, $\mathrm{\langle Fe\rangle}$, Mg b, Ca
4227, C 4668) and one age-related indicator (corrected H$\beta$) for
each of our galaxies that we will analyze.

\section{Mock Catalog Comparison and Correction}
To confirm that we can accurately measure gradients using this annulus 
method, we test the procedure on mock galaxy spectra generated with known 
metallicity gradients. To create this catalog, we use the NASA-Sloan Atlas 
(described in \citealt{blanton11a}) as our reference for the properties of 
similar red and old galaxies. We apply the same selection criteria to the 
NASA-Sloan Atlas as in Section \ref{sec:sampleselect} to ensure we only 
have red galaxies, and then randomly select galaxies from the remaining 
catalog. We assign the Sersic index $n$, the Sersic half-light radius 
(in kpc), the absolute magnitude in the $ugriz$ bands, the stellar mass, 
and the axis ratio $b/a$ for each randomly chosen galaxy; we
then assign a random redshift in our sample ($0.02$ to $0.24$) to the galaxy. 
Based on this new redshift, we rescale the half-light radii into angular
units and the luminosities into fluxes for each galaxy. 
We then perform the identical volume limiting cuts based on redshift and 
V-band magnitude for each galaxy, creating a sample of red galaxies that 
would pass all our real sample’s cuts as well.

We next create mock spectra using Flexible Stellar Population Synthesis code 
(FSPS); \citep{conroy09a, conroy10a}, run via the Python-FSPS modules 
written by Daniel Foreman-Mackey, for a range of metallicities from 
$0.10$ to $-1.00$, spaced at $0.01$ dex, all with identical 10 Gyear ages. 
To properly create the mock spectra we would observe for a galaxy with 
a metallicity gradient, we create a pixel grid (resolution of $0.1$ 
arcsec/pixel) where each pixel’s distance from the center determines its 
metallicity and therefore which FSPS-generated spectrum it is assigned. 
We generate the profile of our mock galaxies given their Sersic indices 
and radii and the axis ratio $b/a$, which lets us know the fraction of the 
luminosity coming from each pixel. We then assign a metallicity to each 
pixel based on its radial distance from the center by using a simple 
linear fit with constant and slope parameters chosen by us. The metallicity
profile follows the light profile of the galaxy --- steeper along the minor
axis as defined by the $b/a$ ratio of the galaxy.

Finally, we must generate a model for the aperture these galaxies are 
observed with. We convolve an image of the SDSS-I and -II fiber aperture 
(3 arcsecond diameter), placed at the exact center of each galaxy, with a 
double-Gaussian PSF. The Gaussians are both wavelength-dependent and 
variable across a range of possible seeings based on the actual BOSS seeing.
For the core Gaussian, we define a mean FWHM of 1.5 arcseconds at 6000 
angstroms, with variability around the mean of 0.3 arcseconds. The second 
Gaussian has a mean FWHM of 5.0 arcseconds at 6000 angstroms, also with
variability around the mean of 0.3 arcseconds. For the wavelength dependence, 
we assume a $\lambda^{-1/5}$ dependence for both Gaussians. The second 
Gaussian integral is weighted with a factor of 0.1 relative to the first.

With our aperture and galaxy image created, we then multiply the two 
together to model how much light at each pixel we would observe, and then 
weight the FSPS-generated spectrum of the metallicity of the pixel by that 
factor, finally summing all the pixels to generate a single mock spectrum 
that represents what we would observe for a galaxy with the metallicity 
gradient we have assigned.

We do not include noise in our procedure. Although noise in the spectra will 
cause noise in the result, it will not change the expectation value of the 
result; i.e. it will not change the expected slope of the measured profile.

\begin{figure*}[t!]
  \centering
  \includegraphics[scale=0.65, angle=90]{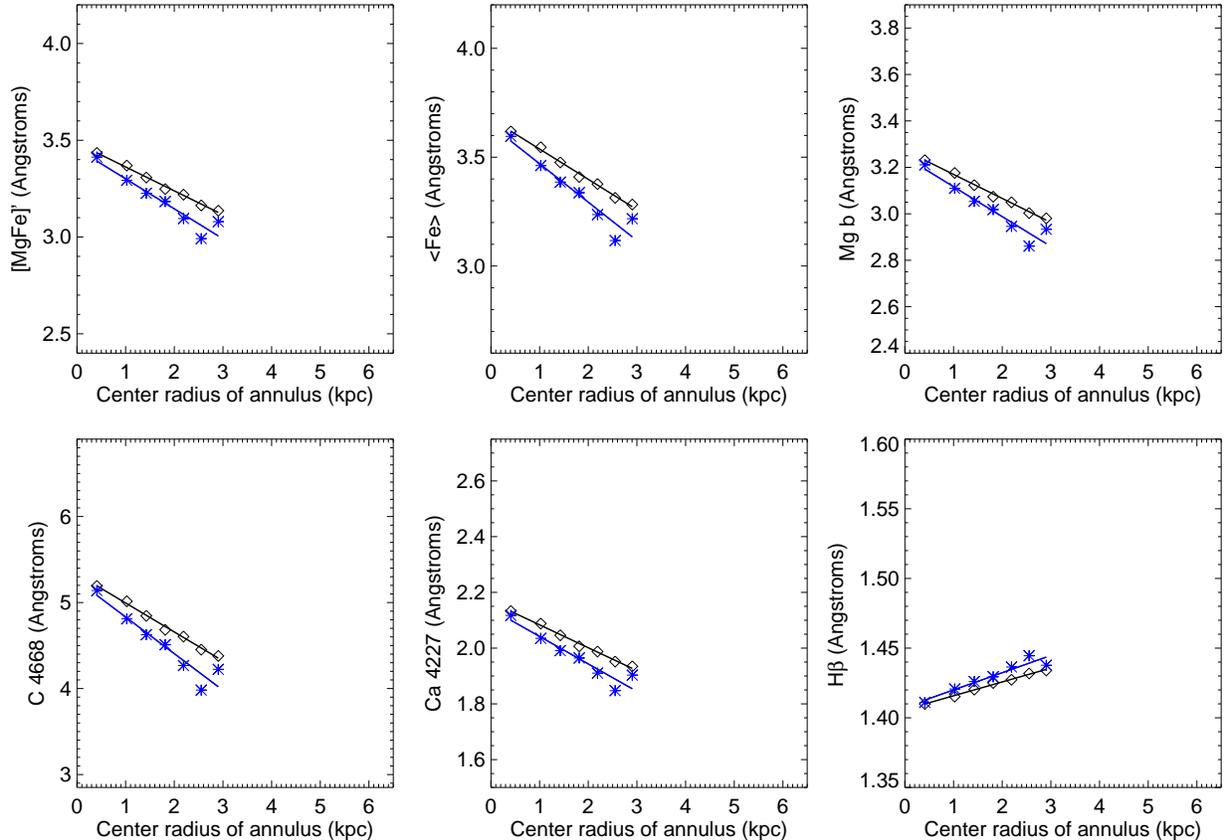}
  \caption{Differences between the gradient input to galaxies and the
gradient we measure using our annulus method for each indicator. Only one 
mass bin ($10.3 < \log(M) < 10.7$) is shown for simplicity, but all mass 
bins were very similar. The input data is shown with black diamonds, 
and the output measured is shown with blue stars. There is a consistent 
steepening of the true gradient, as well as a slight constant offset, 
which we will take into account with a correction factor in our analysis 
and future plots.}
  \label{fig:mockresults}
\end{figure*}

Following this process, we create a mock catalog of 22,000 galaxies with 
known metallicity gradients to analyze identically to our real sample. We 
run it through the same procedures as for the real data to extract the 
metallicity indicators in each annulus for each mass bin and compare to 
the metallicity indicators we input at each radial point. The results of 
this are shown in Fig. \ref{fig:mockresults}. The black lines are the 
gradients in each indicator that were input, and the blue line shows what 
our annulus measuring code returns as outputs. We test a range of input 
metallicity gradients of similar magnitudes to our measured results to 
ensure that our method works for both steep and shallow gradients. We find 
that our method leads to a slight constant offset from our inputs and a 
minor slope steepening (around $15-25\%$, depending on mass bin and 
indicator). However, the changes are not so dramatic as to invalidate the 
method. This steepening of the gradient likely occurs for two reasons: 
first, with an axis ratio not equal to one, our circular annuli are 
overlayed on elliptical constant-metallicity contours (based on the galaxy 
$b/a$ axis ratio). This will cause some data from lower metallicities to 
be included. This lowers the indicator values. Secondly, the PSF we use 
smears light from the entire galaxy image into the aperture, albeit at a 
very low weighting on the low-metallicity edges. This, too, will lower 
the indicator values, but will have a greater impact at high redshift 
(larger effective aperture) due to the differences in how fast the 
PSF falls off away from the center (a Gaussian) and how fast the metallicity 
declines away from the center (log-linearly), resulting in a steepening 
gradient as well as a constant offset.

To examine these differences, we compare the measured Lick index 
values that we input to our recorded output values after the analysis, 
as shown in Fig. \ref{fig:mockresults}.

We find that our results are sensitive to the axis ratio distribution 
and the number of objects in each bin --- a large percentage of objects 
in a bin with extreme axis ratios can cause large (and not real) scatter 
in the two annulus points calculated from that bin. We find it that this
effect is minimized when we work with data sets larger than 20,000 galaxies
for our mock catalogs, and as our real sample is approximately 6 times
larger than that, we are confident that the scatter introduced by our
methodology due to this effect is minor.

As a result of these tests, we can estimate the correction factor in the slope 
needed to make our mock catalog output the same metallicity gradient that 
we input in each indicator. For our presented results below, we will apply 
this small correction factor, which should account for some of these effects
in the real sample.

\section{Data Analysis and Procedures}
\label{sec:dataanalysis}

The SDSS Main sample of galaxies does not have many spectra of
multiple parts of individual galaxies to study how the metallicity
varies as a function of radius in a single object.  Instead, this
sample was entirely observed with single, central fibers with 3 arcsec
diameters. This constant angular aperture means that observing a
galaxy at a low redshift will measure the metallicity in a smaller
physical aperture around the center of the galaxy than a galaxy
observed at higher redshift. This approach follows that of
\cite{yan12a}, who measured emission lines. Here, we are attempting to
localize the absorption lines by stacking galaxies and computing the
values in question in annuli of the galaxies.

Each stellar mass bin is handled separately. Within each stellar mass bin, 
the objects 
are divided up into redshift bins with the goal of having the maximal number 
of bins without having any span too large a redshift range or contain too few 
objects for statistical power. We need to balance the concern of flattening 
the true gradient by having too few bins (see discussion in \citealt{yuan13a}) 
with the concern that too many bins will cause differentials between bins to 
be too small to be observed with our uncertainty being on the same order of 
magnitude as the actual changes. We tested many binning choices and find that 
changes to the binning choices have minimal impact in the determination of the 
gradients, indicating that our results are relatively robust.

Once the stellar mass bins are subdivided into redshift bins, we
compute the average value for the line and continuum luminosity of
each metallicity indicator measured above for all the galaxies falling
into that bin. For details of this sample's binning, see Table
\ref{table:binning}.

We then can compute the Lick index for an annulus by calculating

\begin{equation}
\mathrm{\frac{line~luminosity_{zbin 1} - line~luminosity_{zbin 2}}{continuum~luminosity_{zbin 1} - continuum~luminosity_{zbin 2}}}
\label{eqn:ew_annulus}
\end{equation}
for adjacent redshift bins ``zbin 1'' and ``zbin 2.'' Knowing the
redshift of each object in a redshift bin and the SDSS aperture size
(3 arcseconds diameter), we can then compute the average redshift of
each galaxy in a bin and then convert that to a physical radius
assuming standard cosmology ($H_{0}=70$ km s$^{-1}$ Mpc$^{-1}$,
$\Omega_{M}=0.3$, $\mathrm{\Omega_{\Lambda}}=0.7$). A radius is then
assigned to each annulus by taking the midpoint of the two redshift
bins' physical radii that the annulus is computed from. Thus, we are
left with the pairs of points (Lick index, physical radius) for each
stellar mass bin, which are plotted in Figure \ref{fig:data_gradient}
to measure a gradient. The one exception to this is the innermost
point --- this one is not an annulus but rather just the innermost
redshift bin, and it is assigned a radius of half its maximum
extent. One important detail that results from treating the innermost
bin this way and binning in equal-redshift spacings is that the
innermost redshift bin in each mass bin has a very large amount of
impact on the gradient calculations --- it not only is used twice
(once in the difference between bins 1 and 2 and once by itself as the
innermost point), but it covers the largest range of radius and has
the smallest number of objects of all the redshift bins in that
stellar mass bin.  Finally, because our annuli are no longer fully
independent of each other, there should be nonzero off-diagonal values
in the covariance matrix. We neglect these terms and only consider the
diagonal variance in plotting error bars for each point and in fitting
for gradients.

This analysis is repeated identically, but replacing stellar mass with
velocity dispersion in the initial step. We perform this step mainly
for confirmation that our derived masses are accurate, but also so
that we can more directly compare the results we find to papers that
only report trends with velocity dispersion. Below we will present the
results for both binning schemes, but as will be seen there are only a
small number of differences.

In addition, we also calculate an average galaxy radius for each
stellar mass bin using the average elliptical mass-radius relation
discussed in several papers (\citealt{chiosi12a, shen03a}, among
others). For high-mass ellipticals such as our sample here, the
relation is approximately $\log{R_{1/2}} = 0.54\log{M/M_{*}} -
5.25$. For the stellar mass value, we use the average stellar mass in
each stellar mass bin. As mentioned in those papers, this half-mass
radius is not strictly identical to the half-light, or effective
radius $R_{e}$, but is usually quite close. In this paper we will use
$R_{1/2}$ as a proxy for $R_{e}$ in comparison to other works and
refer to it as $R_{e}$ only from now on. This will allow for better
comparison to theoretical works later in this paper.

\begin{table}[t!]
  \centering
  \caption{Binning of Galaxies in this Paper}
  \begin{tabular}{ c | c | c | c | c | c}
$\log{(M)}$ Range & Objects & $z$ Min & $z$ Max \\
\hline
$10.0 < \log(M) < 10.3$ &  754 &  0.0200 &  0.0317 \\
$10.0 < \log(M) < 10.3$ &  1762 &  0.0317 &  0.0434 \\
$10.0 < \log(M) < 10.3$ &  2695 &  0.0434 &  0.0550 \\
$10.0 < \log(M) < 10.3$ &  3835 &  0.0550 &  0.0667 \\
$10.0 < \log(M) < 10.3$ &  6265 &  0.0667 &  0.0783 \\
$10.0 < \log(M) < 10.3$ &  8135 &  0.0783 &  0.0900 \\
$10.3 < \log(M) < 10.7$ &  1273 &  0.0206 &  0.0348 \\
$10.3 < \log(M) < 10.7$ &  2798 &  0.0348 &  0.0490 \\
$10.3 < \log(M) < 10.7$ &  4509 &  0.0490 &  0.0632 \\
$10.3 < \log(M) < 10.7$ &  8548 &  0.0632 &  0.0774 \\
$10.3 < \log(M) < 10.7$ &  11602 &  0.0774 &  0.0916 \\
$10.3 < \log(M) < 10.7$ &  11929 &  0.0916 &  0.1058 \\
$10.3 < \log(M) < 10.7$ &  16872 &  0.1058 &  0.1200 \\
$10.7 < \log(M) < 11.0$ &  830 &  0.0300 &  0.0514 \\
$10.7 < \log(M) < 11.0$ &  2268 &  0.0514 &  0.0729 \\
$10.7 < \log(M) < 11.0$ &  4442 &  0.0729 &  0.0943 \\
$10.7 < \log(M) < 11.0$ &  6107 &  0.0943 &  0.1157 \\
$10.7 < \log(M) < 11.0$ &  9964 &  0.1157 &  0.1371 \\
$10.7 < \log(M) < 11.0$ &  12070 &  0.1371 &  0.1586 \\
$10.7 < \log(M) < 11.0$ &  15466 &  0.1586 &  0.1800 \\
$11.0 < \log(M) < 11.5$ &  691 &  0.0700 &  0.0983 \\
$11.0 < \log(M) < 11.5$ &  1374 &  0.0983 &  0.1267 \\
$11.0 < \log(M) < 11.5$ &  2594 &  0.1267 &  0.1550 \\
$11.0 < \log(M) < 11.5$ &  4092 &  0.1550 &  0.1833 \\
$11.0 < \log(M) < 11.5$ &  5628 &  0.1833 &  0.2117 \\
$11.0 < \log(M) < 11.5$ &  7346 &  0.2117 &  0.2400 \\
\end{tabular}  
\tablecomments{Details of how galaxies are binned by stellar mass in this 
paper. A bin is 
first divided on the basis of the stellar mass of the galaxies (units are 
solar masses). Following that, the galaxies are divided into bins of equal 
size in redshift. All the limits are set such that the lower limit is 
inclusive and the upper is exclusive to avoid overlap.}
  \label{table:binning}
\end{table}

With the values computed for the various metallicity indicators in
each annulus, we chose to use EZ\_Ages \citep{graves08a} to compute
the actual metallicities as well as several other parameters. EZ\_Ages
uses iterative fitting of the stellar population models described in
\cite{schiavon07a} to determine the best-fit metallicity and age from
the Lick indices we have measured. We applied EZ\_Ages to the stacked
annuli, not the individual objects. We use solar isochrones and a
Salpeter initial mass function with exponent 1.35 as inputs for all
galaxies. Changing to $\alpha$-enhanced isochrones was tested with
minimal impact on the overall results, although a small number of
points no longer fit on the age-metallicity grid. The results of the
EZ\_Ages fitting is shown in Figure \ref{fig:ezages_gradient}. With
metallicities computed, we perform linear regression to compute the
gradient for each stellar mass and velocity dispersion bin separately.

One caveat of the use of EZ\_Ages is that it uses SSP models.
As we are averaging over a range of galaxy types and potential evolution
paths, the already-approximate SSP approach will have more issues.
We thus suggest that care be taken in interpreting the age and metallicity
results too strongly.

\section{Results}

Our results are shown in Figures
\ref{fig:data_gradient}--\ref{fig:ezages_gradient_rre_sigma}
below. The first four (Figures \ref{fig:data_gradient},
\ref{fig:ezages_gradient}, \ref{fig:data_gradient_rre}, and
\ref{fig:ezages_gradient_rre}) show the results as a function of
mass. Figure \ref{fig:data_gradient} displays the Lick indices as a
function of physical annulus radius. Figure \ref{fig:ezages_gradient}
shows the resulting EZ\_Ages model parameters as a function of
physical radius. Figures \ref{fig:data_gradient_rre} and
\ref{fig:ezages_gradient_rre} show the same data, but versus $R/R_e$,
the radius scaled to the galaxy effective radius. The second four
(Figures
\ref{fig:data_gradient_sigma}--\ref{fig:ezages_gradient_rre_sigma})
show the same results as a function of velocity dispersion.

The lines plotted in all eight are the linear fitting results shown in
Table \ref{table:gradients} and Table \ref{table:gradients_re} (for
stellar mass binning) and Table \ref{table:gradients_sigma} and Table
\ref{table:gradients_re_sigma} (for velocity dispersion binning) where
the exact values are listed for more precise comparisons. The units
are dex $\mathrm{kpc^{-1}}$ in radius. In addition, we list the line
intercept at $R=2.5$ kpc and $R/R_e = 0.5$, effectively the mean
aperture-corrected Lick indices and parameters of early-type SDSS
galaxies as a function of mass and velocity dispersion.

Finally, we calculate gradients versus $\log r$ for [Fe/H], [Mg/Fe],
and [C/Fe] for each stellar mass and velocity dispersion bin. These
logarithmic gradients are shown in Figures \ref{fig:massgrad} and
\ref{fig:massgrad_sigma}. The values are given in Tables
\ref{table:loggrads} and \ref{table:loggrads_sigma}. We find in
general very little dependence of the gradients on either mass or
velocity dispersion. The only exceptions are a very small trend
towards flattening gradients in [Fe/H] as a function of velocity
dispersion or mass, with the exception of the lowest velocity 
dispersion or mass bin gradient of [Fe/H] which is much steeper than the 
others.

Note that in the above results we found a systematic mis-measurement
for one annulus of the H $\beta$ line in the $10.7 < \log{M} < 11.0$
mass bin, due to part of the redshift range it covers.  The H $\beta$
line for one bin is redshifted on top of the OI 5577 \AA\ sky line;
errors in the sky subtraction at that location lead to an abnormally
low flux measured in the line. This leads to H$\beta$ absorption that
appears too large, and an age that is too young, as well as a slight
increase in [Fe/H] and [Mg/Fe] for that one point. We retain this data
point on all plots, but do not use it in the fits; it is tinted a
slightly lighter red in color to indicate the presence of bad
data. This issue is not as clearly present in the velocity dispersion
binning, likely due to its wider redshift coverage per bin, which
dilutes the effect of the error; no masking is used in this case.

\begin{figure*}[t!]
  \centering
  \includegraphics[scale=0.65, angle=90]{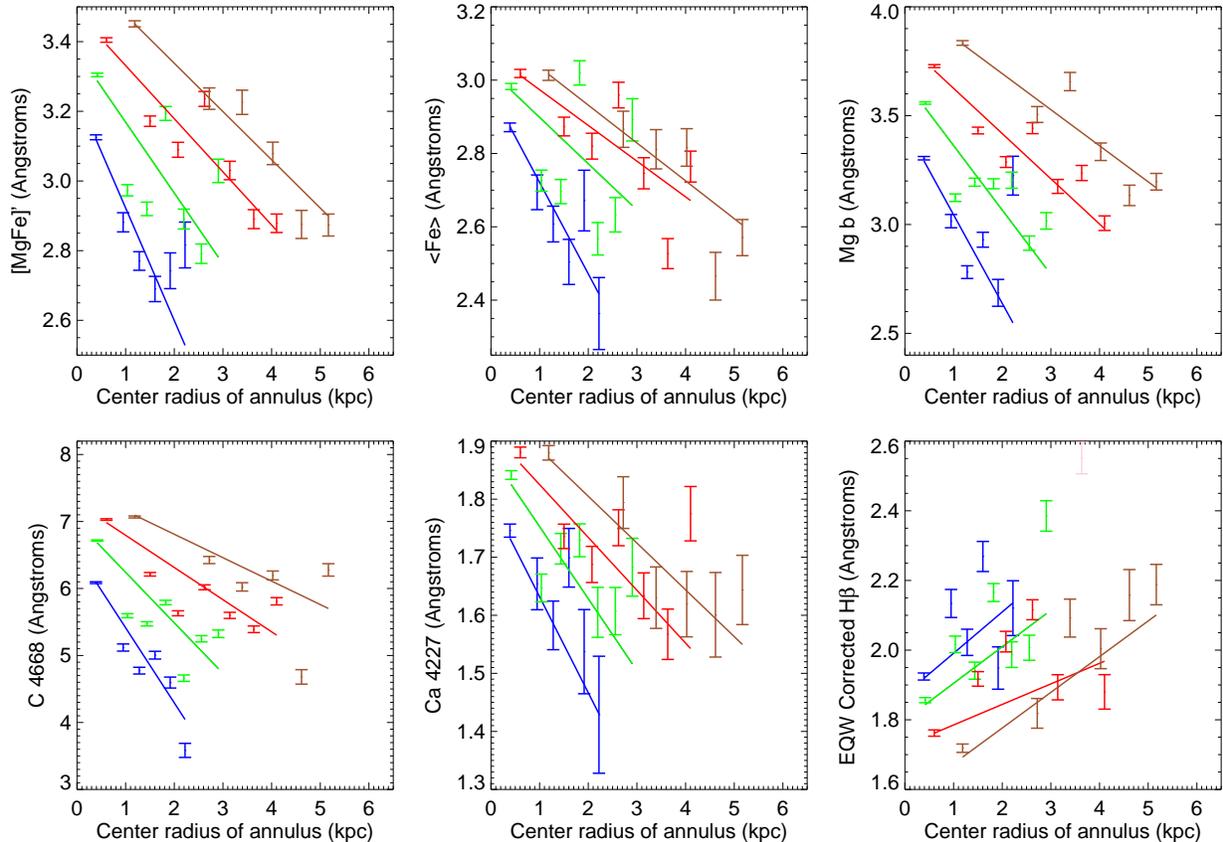}
  \caption{The gradients found for each metallicity indicator. Each stellar mass bin 
is represented with a different color: $10.0 < \log{M} < 10.3$ is blue, 
$10.3 < \log{M} < 10.7$ is green, $10.7 < \log{M} < 11.0$ is red, and 
$11.0 < \log{M} < 11.5$ is brown}
  \label{fig:data_gradient}
\end{figure*}

\begin{figure}[t!]
  \centering
  \includegraphics[scale=0.45, trim=0 2.7in 0 0, clip=true]{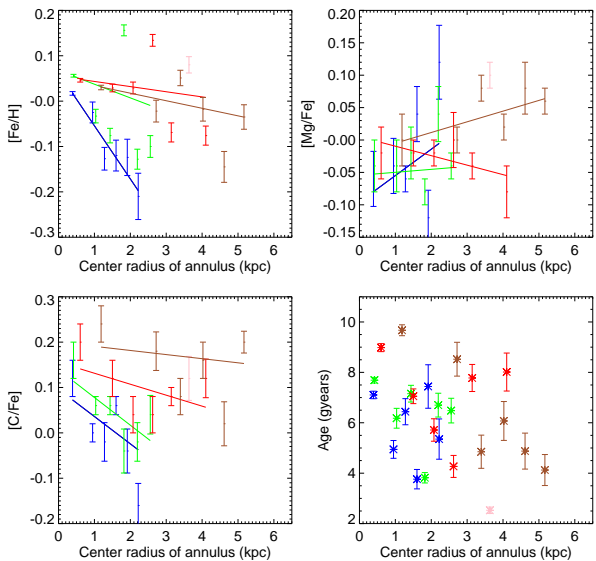}
  \caption{The gradients found for the metallicity, age, and several 
[$\alpha$/Fe] ratios computed using EZ\_Ages. Colors are the same as
Figure \ref{fig:data_gradient}. Binning is done based on stellar mass.}
  \label{fig:ezages_gradient}
\end{figure}

\begin{figure*}[t!]
  \centering
  \includegraphics[scale=0.65, angle=90]{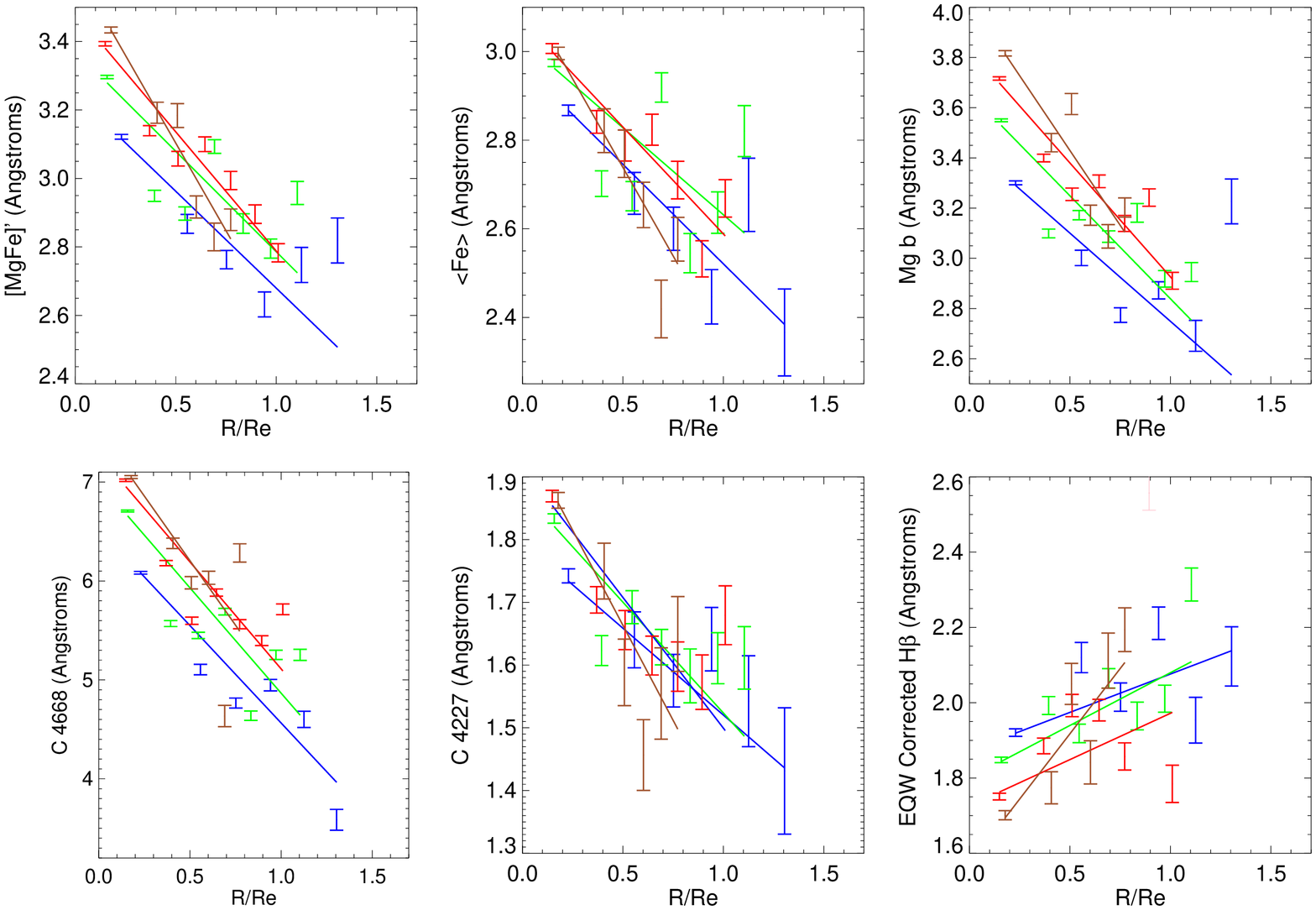}
  \caption{Similar to Figure \ref{fig:data_gradient}, but with all radii 
scaled to the effective radius. Colors are the same as Figure 
\ref{fig:data_gradient}. Binning is done based on stellar mass.}
  \label{fig:data_gradient_rre}
\end{figure*}

\begin{figure}[t!]
  \centering
  \includegraphics[scale=0.45, trim=0 2.7in 0 0, clip=true]{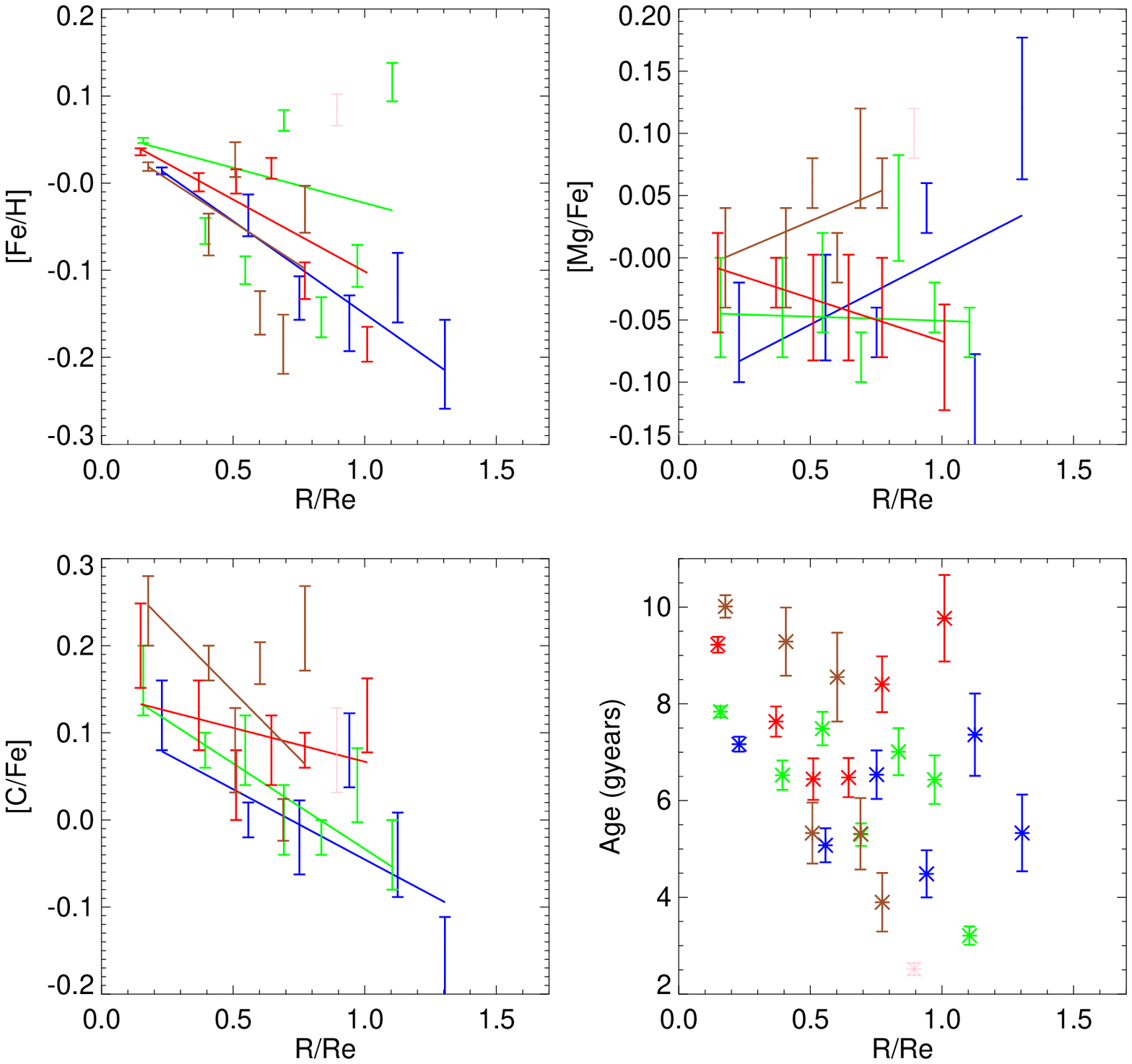}
  \caption{Similar to Figure \ref{fig:ezages_gradient}, but with all radii 
scaled to the effective radius. Colors are the same as Figure 
\ref{fig:data_gradient}. Binning is done based on stellar mass.}
  \label{fig:ezages_gradient_rre}
\end{figure}

\begin{figure*}[t!]
  \centering
  \includegraphics[scale=0.65, angle=90]{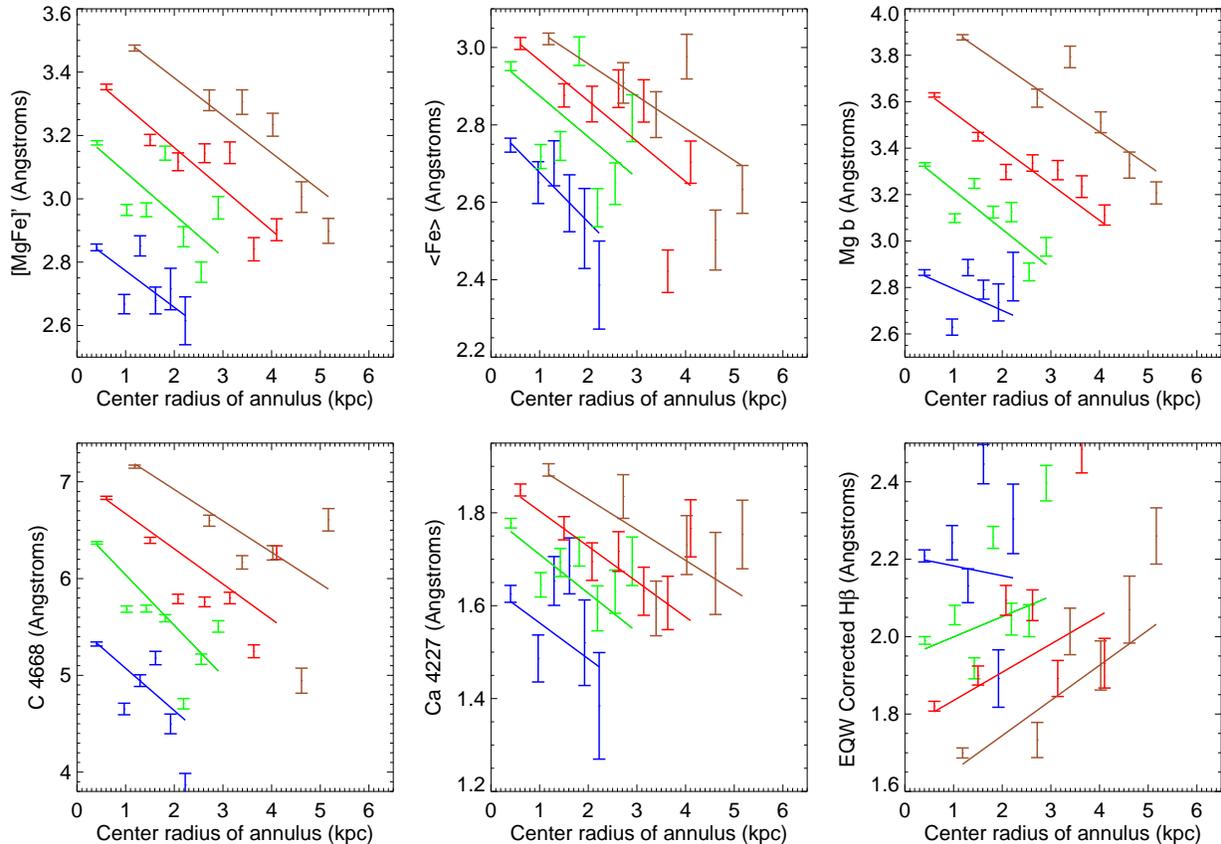}
  \caption{The gradients found for each metallicity indicator, but with the 
initial binning done by velocity dispersion and not stellar mass. Each velocity
dispersion bin is a different color: 30 km s$^{-1}$ to 125 km s$^{-1}$ blue, 
125 km s$^{-1}$  to 185 km s$^{-1}$ green, 185 km s$^{-1}$ to 230 km s$^{-1}$ 
red, 230 km s$^{-1}$ to 325 km s$^{-1}$ brown.}
  \label{fig:data_gradient_sigma}
\end{figure*}

\begin{figure}[t!]
  \centering
  \includegraphics[scale=0.45, trim=0 2.7in 0 0, clip=true]{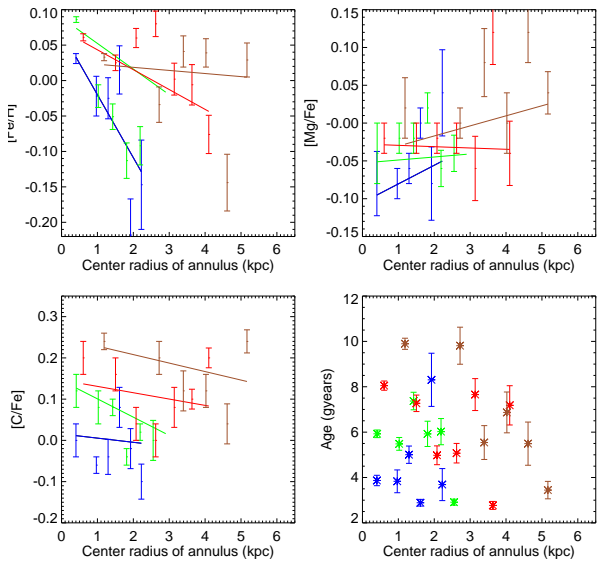}
  \caption{The gradients found for the metallicity, age, and several 
[$\alpha$/Fe] ratios computed using EZ\_Ages. Colors are the same as Figure 
\ref{fig:data_gradient_sigma}. and binning is done on the basis of velocity 
dispersion.}
  \label{fig:ezages_gradient_sigma}
\end{figure}

\begin{figure*}[t!]
  \centering
  \includegraphics[scale=0.65, angle=90]{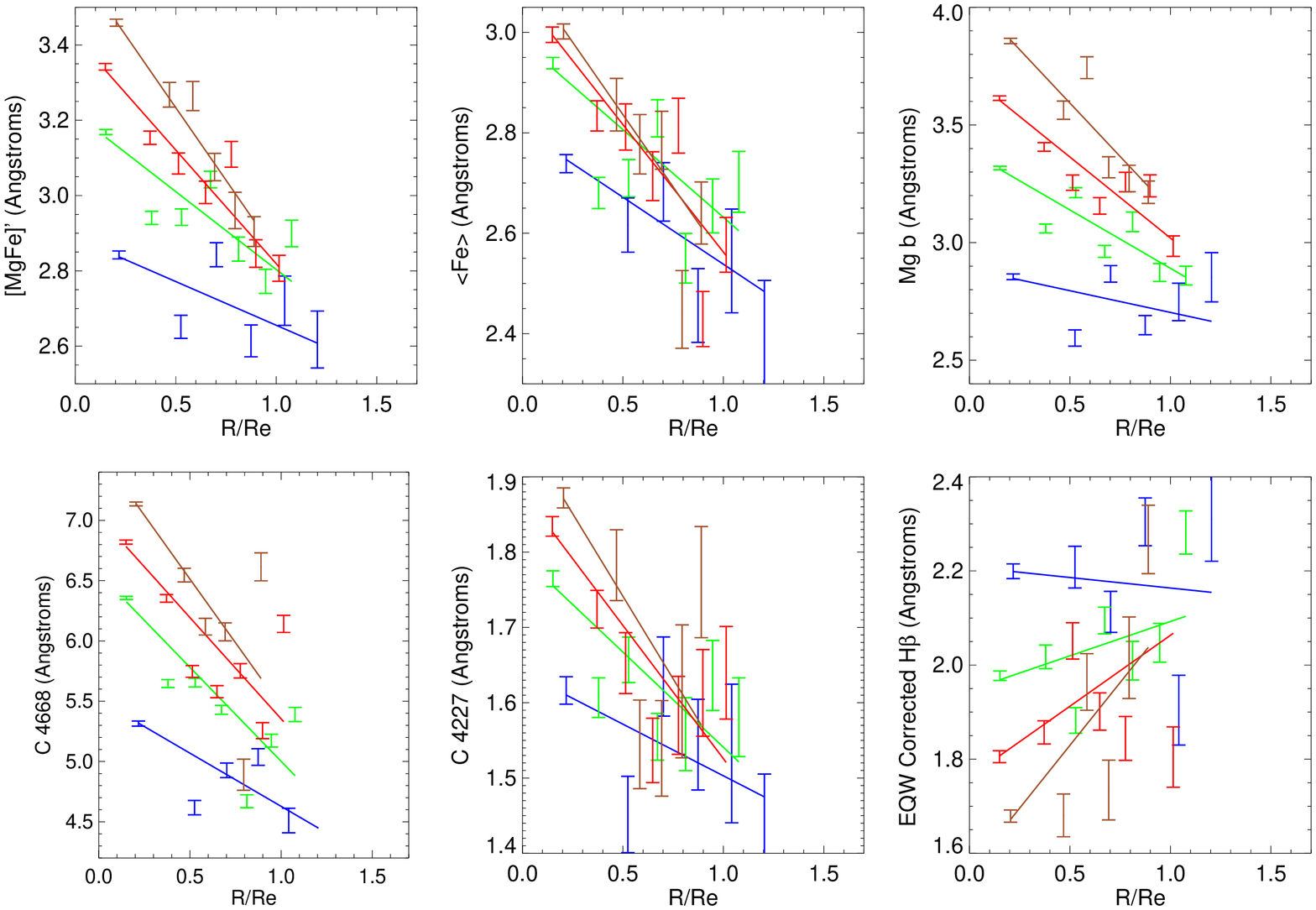}
  \caption{Similar to Figure \ref{fig:data_gradient_sigma}, but with all 
radii scaled to the effective radius. Colors are the same as Figure 
\ref{fig:data_gradient_sigma}. Binning is done based on velocity dispersion.}
  \label{fig:data_gradient_rre_sigma}
\end{figure*}

\begin{figure}[t!]
  \centering
  \includegraphics[scale=0.45, trim=0 2.7in 0 0, clip=true]{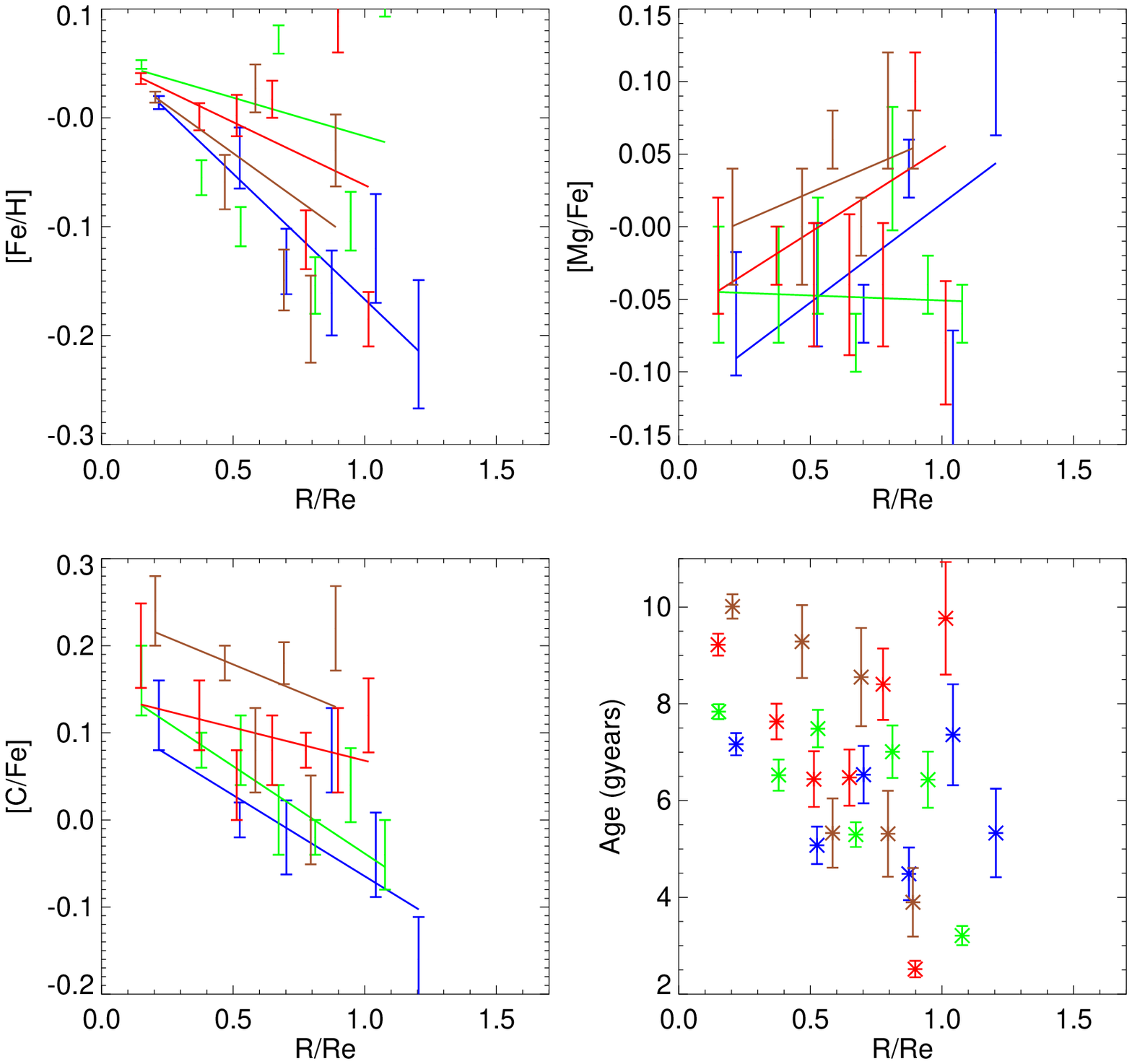}
  \caption{Similar to Figure \ref{fig:ezages_gradient_sigma}, but with 
all radii scaled to the effective radius. Colors are the same as Figure 
\ref{fig:data_gradient_sigma}. Binning is done based on velocity dispersion.}
  \label{fig:ezages_gradient_rre_sigma}
\end{figure}

There are a wide range of important conclusions to draw from these
data. Before moving to discuss them in the context of previous
observations and theory, it's worth simply listing some of the notable
facts.
\begin{enumerate}
    \item All the metal-related Lick indices show a signficantly
      negative gradient when plotted both versus physical radius and
      $R/R_{e}$.
    \item Gradients for stellar population parameters [Fe/H] and
      [C/Fe] are generally declining, while the gradient for [Mg/Fe]
      is generally flat.
    \item The stellar population ages have very large scatter, with
      ages ranging from 3 to 12 billion years.
    \item There is a slight increase in the central value of [C/Fe]
      with mass, while the central [Fe/H] and [Mg/Fe] values are
      nearly constant (where ``central'' corresponds to the inner 0.5
      kpc).
    \item The logarithmic gradients have little detectable dependence
      on mass or velocity dispersion, with the greatest dependence
      being a slight flattening of the [Fe/H] gradient as those
      parameters increase.
    \item Most of the dependence of the gradient with physical radius
      on mass is accounted for by the change in $R_e$ with mass.
    \item There are no dramatic differences between the results from
      the stellar mass binning and the velocity dispersion binning.
\end{enumerate}

We can determine metallicities rather well, but due to the uncertainty
in our H$\beta$ measurements and corrections, our age 
determinations have a large
amount of scatter. Some of the H$\beta$ uncertainty is driven by the
inherent scatter in the [OIII] correction to H$\beta$ line infill,
some results from the extra noise in measuring two lines (instead
of one), and some results from the expected slight
differences amongst these galaxies that will add some scatter to our
points as well. In addition to this, the redshift range that a 
single data point covers will include some real age differences of the 
galaxies due to the universe age at that point, which is not accurate
if we wish to infer age gradients in an actual galaxy. This effect is 
at most 2 Gyears across a single bin, though. Due to the lack 
of trend we find in the age points and the likelihood of these
gradients not measuring what we wish to measure, we
do not fit to these data points --- they are simply shown with
error bars at each radius.

We can also see that our stellar mass bins and velocity dispersion
bins show similar trends, as we would expect. If we consider the
$R/R_{e}$ plots (Figures \ref{fig:data_gradient_rre},
\ref{fig:ezages_gradient_rre}, \ref{fig:data_gradient_rre_sigma},
\ref{fig:ezages_gradient_rre_sigma}; values given in Table
\ref{table:gradients_re} and \ref{table:gradients_re_sigma}), we also
find that stellar mass or velocity dispersion only influences the
fitted gradients slightly --- the [Fe/H] plot shows a steadily
increasingly negative central metallicity as stellar mass increases,
but the gradient stays mostly constant (contrary to expectations,
although similar to recent findings in \citealt{pastorello14a}). This
trend in central metallicities is loosely reflected in the
$\mathrm{[MgFe]'}$ and $\mathrm{\langle Fe\rangle}$ plots as well, as
would be expected. Other metals such as [C/Fe] and the indicators C
4668 and Mg b show no clear dependencies on stellar mass in gradient
or central values, however.

Below, we will use both the mass and the velocity dispersion results
for comparison to prior works, depending on what data the paper in
question offers, because our qualitative conclusions are the same
regardless.

\begin{figure}[t!]
  \centering
  \includegraphics[scale=0.45, trim=0 2.7in 0 0, clip=true]{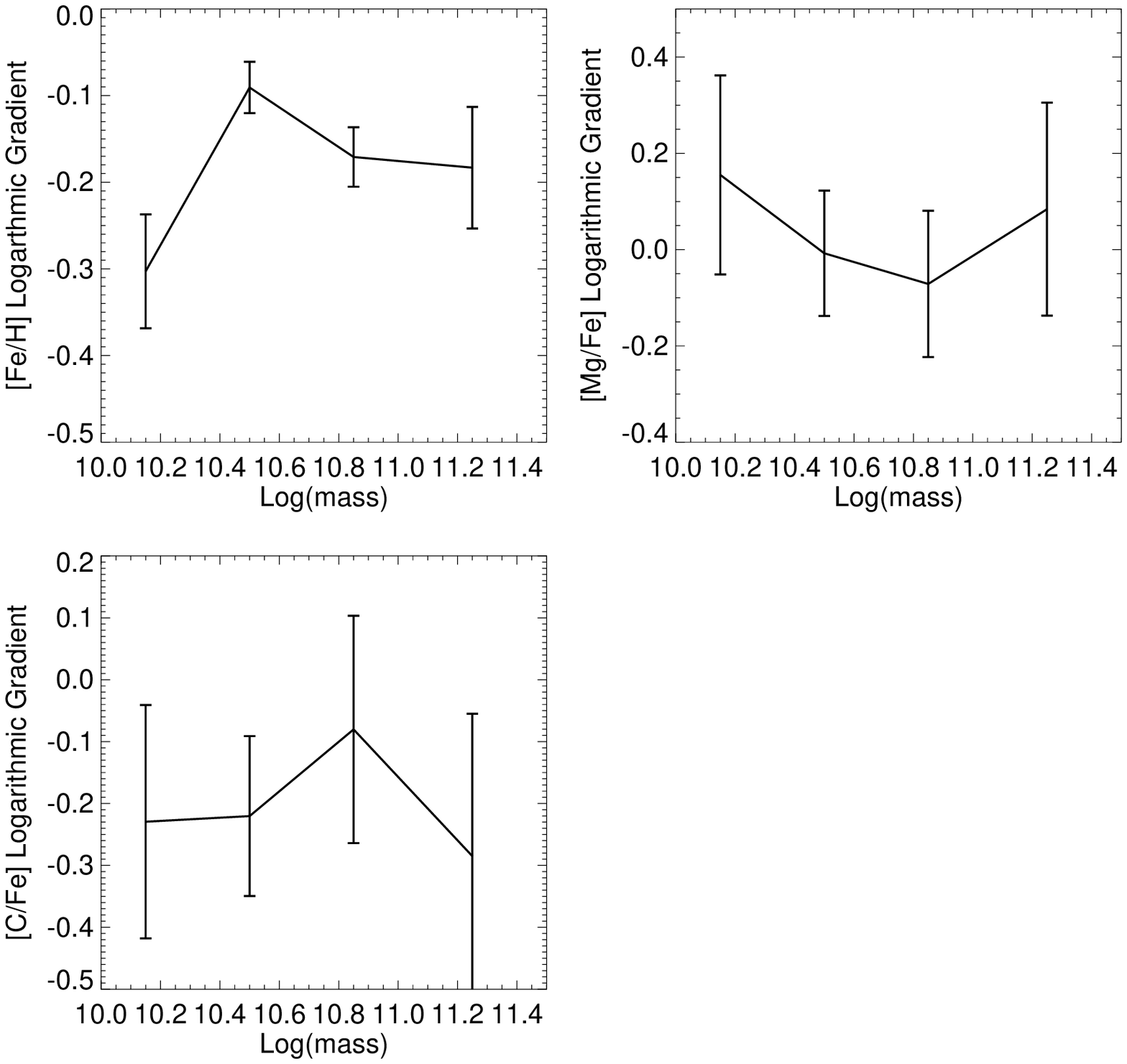}
  \caption{The stellar mass dependence of the [Fe/H], [Mg/Fe], and
    [C/Fe] logarithmic gradients as calculated in Table
    \ref{table:loggrads}. Broadly speaking we find a constant [Fe/H]
    (with the exception of the lowest stellar mass bin which shows a
    much steeper gradient), [Mg/Fe] and [C/Fe] gradient as a function
    of stellar mass.}
  \label{fig:massgrad}
\end{figure}

\begin{figure}[t!]
  \centering
  \includegraphics[scale=0.45, trim=0 2.7in 0 0, clip=true]{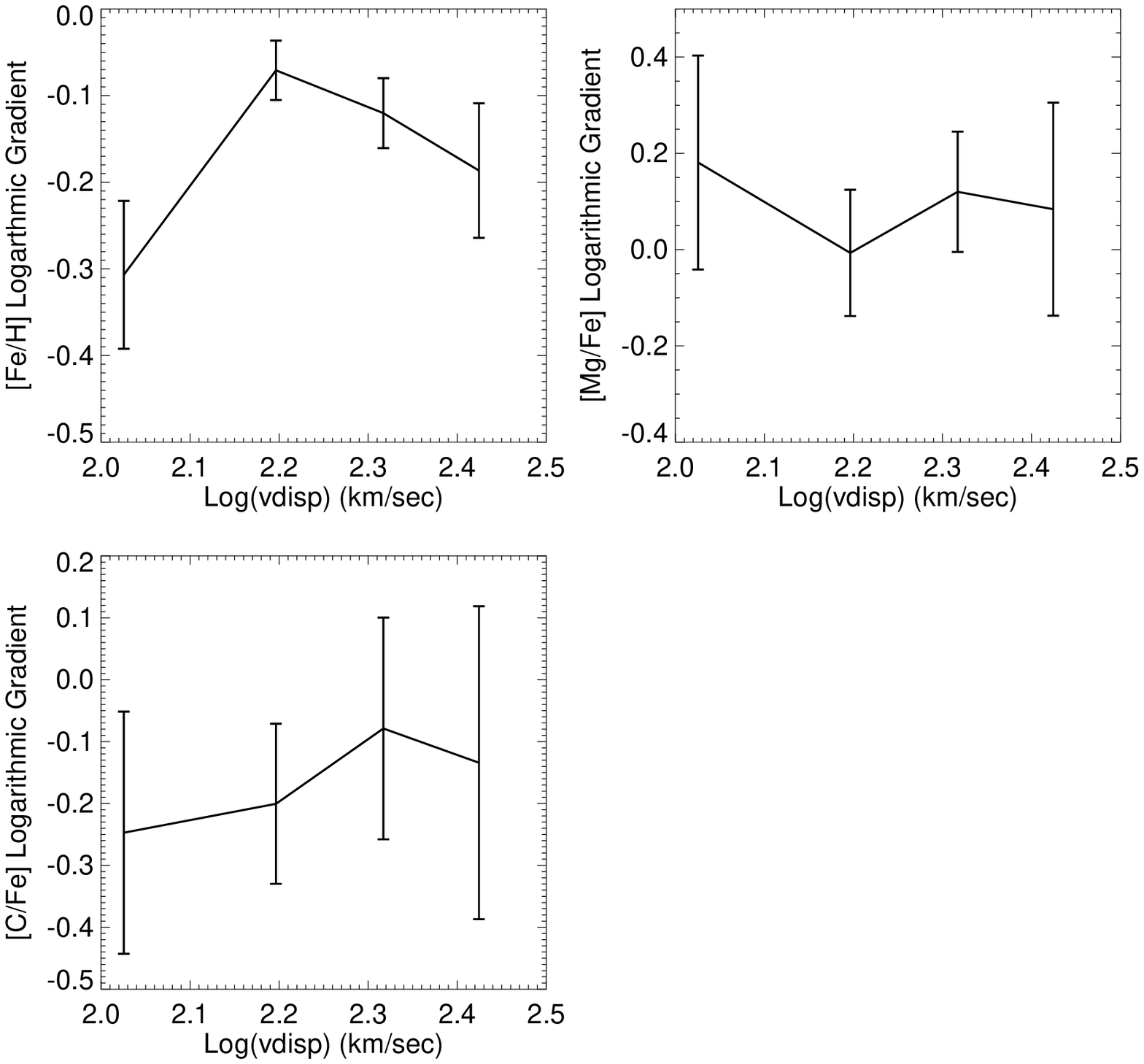}
  \caption{The velocity dispersion dependence of the [Fe/H], [Mg/Fe],
    and [C/Fe] logarithmic gradients as calculated in Table
    \ref{table:loggrads_sigma}. Broadly speaking we find a very
    slightly flattening [Fe/H] gradient, and a constant [Mg/Fe] and
    [C/Fe] gradient as a function of velocity dispersion.}
  \label{fig:massgrad_sigma}
\end{figure}

\section{Comparison to Previous Observational Studies}
\label{sec:obs}

We have three main results: gradients for the EZ\_AGES outputs
([Fe/H], [C/Fe], and age), the trend in their central values with
mass, and the trend in their gradients with mass. Each of these have
been studied in other papers and we compare with those results here.

Some of the papers we will examine do not use the same stellar
population models --- they instead use the slightly different models
from \cite{thomas03a} (TMB).  These models differ slightly in the
outputs they produce, with the TMB models giving [Z/H] and
[$\alpha$/Fe] instead of the EZ\_AGES output of [Fe/H], [Mg/Fe], and
[C/Fe].  \cite{thomas03a} proposes a conversion between the two
measures of [Fe/H]$=$[Z/H]$- 0.94$[$\alpha$/Fe], which we will use to
compare our observations to those papers that use the TMB models. For
the $\alpha$-elements, the TMB models assume that all
$\alpha$-elements track Mg, so we can compare our [Mg/Fe] to the
[$\alpha$/Fe] values. [C/Fe] will have no analog in those cases.

\subsection{Gradients}

Two large studies integral field studies of the metallicity gradients
in inner regions of galaxies cover well the same radial range we
cover: \cite{rawle10a} and \cite{kuntschner10a}. Both these studies
measure [Z/H] values out to $1 R_{e}$.  Both only calculate a
[$\alpha$/Fe] ratio (which tracks Mg) and not a [C/Fe] ratio. When applying
the [Z/H] to [Fe/H] conversion to match our results, both
studies find a slightly negative logarithmic metallicity gradient (however,
note that the [Fe/H gradient in \cite{rawle10a} is consistent with zero),
matching our observations --- \cite{rawle10a} finds an average
gradient of $-0.05 \pm 0.05$ $dex^{-1}$ and \cite{kuntschner10a} 
finds an average
gradient of $-0.25 \pm 0.11$ $dex^{-1}$, with values ranging from $-0.1$ to
$-0.5$ much like our results. Also in agreement with
our observations, \cite{kuntschner10a} report an [$\alpha$/Fe]
gradient consistent with zero. \cite{rawle10a} finds a very slightly negative
[$\alpha$/Fe] gradient of $-0.06 \pm 0.03$ $dex^{-1}$, which indicates 
a slight trend 
unlike our data, but isn't outside of our error bar range.

\cite{mehlert03a} study the metallicity gradients of early-type
galaxies in the Coma Cluster out to about $R_{e}$, also similar to our
measurements, using the data from \cite{mehlert00a}. Rather than using
EZ\_Ages to model the metallicity and age, they use the models of 
\cite{thomas03a}, with
the differences discussed above.  The results in \cite{mehlert03a}
show declining [Z/H] and constant age and [$\alpha$/Fe] gradient. We
can calcuate similar logarithmic index gradients (per Equation 5 in
\citealt{mehlert03a}, given in Table \ref{table:loggrads}) to compare
the values of the metallicity indicators in our stacked annuli to
their results.  Their [Z/H] gradient matches our [Fe/H] gradient well,
with a gradient of around $-0.1$ to $-0.2$.

\cite{spolaor10a} cover a slightly larger range of radii, out to
between 1 and 3 $R_{e}$.  They analyzed 14 low-mass ellipticals in the
Fornax and Virgo clusters and compared them to several higher mass
ellipticals from previous studies.  They interpret the stellar
populations using the models of \cite{thomas03a}. They find a
declining [Z/H] ($-0.22\pm0.14$ averaged across all mass bins) and
roughly constant [$\alpha$/Fe] gradient, in agreement with our
results.

\cite{greene13a} study gradients in 33 nearby galaxies, also out to a
much larger radius of the galaxies than we do, about 14 kpc.  They
measure similar metallicity indicators to our survey, also using
EZ\_Ages to convert the data into physical parameters.  They find a
[Fe/H] gradient of around $-0.3$ to $-0.5$ dex $\mathrm{kpc^{-1}}$,
significantly steeper than ours (Note that 
here they report gradients versus radius
rather than versus logarithmic radius). This discrepancy may indicate
that the metallicity gradients steepen in the outer parts of galaxies.
Their findings for [Mg/Fe] and [C/Fe] are similar to ours, with a
constant [Mg/Fe] gradient in both velocity dispersion bins like all
four of ours, and a [C/Fe] gradient of around $-0.1$ to $-0.2$ dex
$\mathrm{kpc^{-1}}$ as our data shows as well.

Because the [$\alpha$/Fe] gradients are relatively flat in all cases
aside from that of \cite{rawle10a}, it is reasonable that our [Fe/H]
results and the [Z/H] results from these studies are in agreement.
 
\subsection{Trend in central values with mass}

We next turn to trends with mass in the stellar population parameters
at the centers of galaxies.  Our centermost bins are $R\sim 0.2 R_{e}$
or $\sim$ 0.5 kpc. We find only weak (at best) trends with mass. Central 
[Fe/H] shows a dependence that is not monotonic and is $<0.1$ dex across 
our whole sample. Central [Mg/Fe] shows an small increase  with velocity 
dispersion that is $\sim 0.1$ dex across our whole range of masses or 
velocity dispersion, but the same trend does not exist for 
mass binning

\cite{kuntschner10a} reports their central values at $R_{e}/8$, slightly more
internal than ours but comparable. They too find no monotonic dependence
of the central metallicities on velocity disperson or mass, although their 
results have more scatter (ranging from $0.2$ to around $-0.4$) after we apply 
the [Z/H] to [Fe/H] conversion. Again similarly, they find that the central 
[$\alpha$/Fe] does increase with mass or velocity dispersion, although 
there is a constant offset of about $0.2$ dex higher in their central values 
than we find, part of which can be explained by the more central 
location. Their trend is much clearer than ours is as well.

Turning to \cite{rawle10a}, we find central values recorded at $R_{e}/3$, 
again comparable to our central annulus radius. As with \cite{kuntschner10a}, 
after converting their reported [Z/H] values to [Fe/H], there is no clear 
trend with central velocity dispersion and a larger scatter than we report. 
Their [$\alpha$/Fe] central trend with central velocity dispersion
also matches ours with an increase as velocity dispersion increases, but 
as with \cite{kuntschner10a}, they also find systematically higher values 
for the central [$\alpha$/Fe] than we record and a clearer 
indication of the trend than we see.

\cite{spolaor10a} detect an increase with mass of central [Z/H] and
[$\alpha$/Fe] at radii similar to the smallest radii we probe
($R_{e}/8$); between 100 and 300 km s$^{-1}$ they detect an
increase in [Z/H] of 0.3 dex and in [$\alpha$/Fe] of 0.2 dex. When applying 
the conversion from [Z/H] to [Fe/H], this is close to our results, although 
a slight central $[Fe/H]$ increase as a function of velocity still remains 
that we do not detect and the dependence of [$\alpha$/Fe] is stronger than 
what we find, matching more closely \cite{kuntschner10a} and \cite{rawle10a}.

Using data with similar radial coverage, \cite{mehlert03a} found
similar dependences. These results indicate stronger mass dependence
than we find by about 0.1 dex over this range for
[$\alpha$/Fe]. Assuming the conversion about between [Z/H] and [Fe/H]
it also implies a stronger mass dependence in [Fe/H] than we find,
also by about 0.1 dex.

The galaxies in \cite{greene13a} also show an increase in central
metallicity with mass at a fixed radius, but their centermost bin is
around 2--3 kpc. Our galaxies show a similar trend at that physical
radius.  The trends with mass agree with those in \cite{greene13a} in
the [C/Fe] and [Mg/Fe] ratio, with the former showing slight increase
at all our radii with stellar mass, and the latter being roughly
constant.  This trend with mass persists across all radii in our
survey (with the exception of the lowest mass or velocity dispersion
bin).

\subsection{Trend in gradients with mass}

Figures \ref{fig:massgrad} and \ref{fig:massgrad_sigma} show the
logarithmic gradients as a function of stellar mass and 
velocity dispersion.  Our metallicity gradients are close to constant
or perhaps slightly steepening 
within the errors as stellar mass increases with the only exception 
being the lowest stellar mass and velocity dispersion bins. In the higher 
mass bins, the [Fe/H] gradient anti-correlates with both the 
stellar mass and the velocity dispersion; however, both trends are
very low significance.  For [Mg/Fe] and [C/Fe], we find no significant
trends in the gradients as a function of mass or velocity dispersion.

Both \cite{rawle08a} and its followup paper, \cite{rawle10a}, as well 
as \cite{kuntschner10a} find a large amount of
scatter in gradients.  \cite{kuntschner10a} note a slight trend that
is partially reflected in our data. At low masses,
\cite{kuntschner10a} reports an increasingly negative metallicity
gradient (up to about $3.5 \times 10^{10}$ solar masses), followed by
a flattening metallicity gradient as mass increases further. We have
far fewer points to use to determine trends with mass, but our
lowest-mass point is indeed the steepest gradient, with all
higher-mass ones showing some evidence of flattening out (see Figures
\ref{fig:massgrad} and \ref{fig:massgrad_sigma}, and Tables
\ref{table:loggrads} and \ref{table:loggrads_sigma}).  However the
granularity of our data prevents us from checking in detail at which
mass we find this turnover to be at to further compare to the
observations in \cite{kuntschner10a}.

\cite{spolaor10a} also covers enough of a stellar mass range to check
for any mass dependency of these gradients, and does find a slight
flattening of the metallicity gradient as stellar mass increases for
high-mass ellipticals, although the scatter increases significantly as well. 
The bins that overlap most completely (covering a range of velocity
dispersions $2.1 < \log{\sigma} < 2.6$) have very similar measured
gradients as a function of R/R$_{e}$ ($-0.2$ to $-0.4$ in 
\cite{spolaor10a} and $-0.15$ to $-0.25$ in our sample), although the
trend is reversed.

The work of \cite{ogando06a} provides another reference on metallicity 
gradients for us to compare to. \cite{ogando06a} measure Mg$_{2}$ and H$\beta$ 
only to supply to the TMB stellar models, with a forced [$\alpha$/Fe] of 
$0.3$, slightly higher than our model fits would predict, but within most 
of the error bars. With that, they find [Fe/H] logarithmic gradients 
(converted from the [Z/H] gradients in the paper) ranging from $-0.1$ 
to $-1.2$, but with the bulk of the galaxies found to have gradients 
between $-0.3$ and $-0.8$. They record these as functions of galaxy 
stellar velocity dispersion and mass, and find a slight trend to a 
steepening [Fe/H] gradient as mass or velocity dispersion increases, 
matching our 3 highest mass bins; however, the scatter is large. This is 
in agreement with the results in \cite{kuntschner10a} and our results, 
although again, the observed steepening with increased mass is paired 
with an increase in scatter as well.

Finally, at the larger radii that \cite{greene13a} measure, they find
qualitatively similar trends, with roughly constant gradients in all
indicators as a function of mass.

As will be discussed more in Section \ref{sec:theory}, the scatter observed
in the gradients could be due to the wide variety of merger histories
available to high-mass ellipticals \citep{dimatteo09a}, leading to a
wide variety of potential final, observed gradients for individual
galaxies.

Realistically, the large error bars present in the determination of
the metallicity gradients prevent us from concluding anything too
firm; we mostly can say that it appears that the gradient very
slightly flattens with increasing stellar mass or velocity 
dispersion, and isclose to constant or slightly declining 
above a certain stellar mass or velocity dispersion cutoff around 
$3 \times 10^{10}$ solar masses, which is not in disagreement with 
any of the studies we found.

\subsection{General conclusions}

There is good agreement with finding a negative [Fe/H] gradient, a
constant [Mg/Fe], and a very slightly negative [C/Fe] gradient, even
if the numerical results presented have a fairly large amount of
scatter. For trends in central values as mass (or velocity dispersion)
increases, we find general agreement with an increasing central [Mg/Fe] 
(although our dependence
is less than other compared studies by about 0.1 dex) and constant central
[Fe/H], and increasing central [C/Fe] (where central in this data set
means within $\sim 0.5$ kpc).

Finally, we find that in general we agree with the most of the reported 
trends in gradients of [Fe/H] with mass as well --- with the most negative 
gradient in the lowest
mass objects, and then a flatter although still negative gradient for
higher mass galaxies, with the gradient leveling out to be roughly
constant with mass or perhaps very slightly steepening. [Mg/Fe] 
shows agreement here as well, with no real trends reported with mass in 
the gradients.

Because it is a statistical average, our data set is not sensitive to
the increase in scatter with mass found by \cite{kuntschner10a} and
others in the [Fe/H] gradient.

\section{Comparison to Theory}
\label{sec:theory}

According to a number of theoretical investigations, mergers tend to
flatten gradients and monolithic collapse models tend to steepen them
\citep{pipino10a, dimatteo09a}. Because the detailed history of each
galaxy's growth involves some features of both monolithic collapse and
hierarchical merging models, the predicted results lie along a
continuum of flat to steep gradients, with substantial scatter amongst
the results due to differing degrees of these two effects for each
individual galaxy \citep{pipino10a}. Thus we do not expect perfect
agreement with any individual models but rather that our data lie
somewhere in between results reported by merger-focused simulations
and monolithic collapse-focused simulations.

When comparing our gradient values to those in \cite{hopkins09a}, who
simulated merger models, we find general agreement. Most of their
metallicity gradients are between $-0.1$ and $-0.6$, while our results
(shown in Table \ref{table:loggrads}) are around $-0.1$ to $-0.4$
dex. Similar conclusions are drawn by \cite{kobayashi03a} who finds
gradients in the range of $-0.2$ to $-0.8$. \cite{kobayashi03a}
simulates both merging and monolithically collapsing galaxies, and it
is important to note that within the spread of the results, our data
agrees with both sets. Because there is likely to be natural variation
among galaxies in our sample, we do not rule out that some galaxies
have the steep gradients predicted by models.  \cite{dimatteo09a}
handles mergers in more detail than most simulations, measuring
gradients that result from various mass ratios and various initial
metallicity gradients of merging galaxies. The results again show
metallicity gradients of about $-0.1$ to $-0.4$, but are dependent on
the type of mergers that have occurred to form the final galaxies as
expected.  For reasonable initial metallicity gradients and merger
histories, our results appear in agreement with the simulations of
\cite{dimatteo09a}.

Another interesting property to investigate is the stellar mass
trend. Our results (see Figure \ref{fig:massgrad}) suggest that all
three gradients are roughly constant with stellar mass, with perhaps a
slightly flattening [Fe/H] depending on how much the lowest stellar
mass bin is to be believed. Simulations give varied results depending
on the merger history of the galaxies, with \cite{dimatteo09a} noting
processes that can give rise to both flatter and steeper gradients
depending on the stellar masses of merging galaxies and their initial
gradients, and thus proposing that little trend should exist overall
but scatter should increase. Other models predict no clear trend and
increasing scatter as well (see \citealt{pipino10a,
  kobayashi03a}). Some models do predict a correlation
\citep{kawata03a}, but only take into account monolithic collapse and
not the interplay of mergers as well. In general, though, the
conclusions are broadly suggestive that metallicity gradients should
slightly steepen with increasing stellar mass, albeit with increased
scatter as well \citep{ogando05a} due to higher stellar mass galaxies
being allowed a larger variety of possible evolution paths that will
change how their gradients develop. We find no clear trend in [Fe/H]
gradient with stellar mass or velocity dispersion, with only a weakly
detected flattening.  The error bars in our results are largely driven
by statistical considerations based on the number of objects in each
stellar mass bin, so we are unable to detect any potential intrinsic
scatter in the data that may exist due to our method.

The potential for many types of mergers and the differing gradients
that result from the different stellar mass ratios and initial
metallicity gradients of the progenitors as discussed in detail in
\cite{dimatteo09a} make it difficult to conclude definitively if our
observations agree with theory as simply adjusting the progenitor
properties within reasonable values can change the model predictions
dramatically. To make a more detailed comparison of our results to
theory would require modeling expected merger rates and stellar mass
ratios.  However, for now, we can conclude that our results do not
indicate any clear disagreements --- the gradients we find are within
the ranges predicted by models that incorporate both monolithic
collapse and mergers of many types into the evolution of elliptical
galaxies, and our gradients follow a generally observed trend with
stellar mass.

\section{Conclusions}

The approach used here has drawbacks and challenges but also significant 
advantages relative to previous studies. We can determine only 
mean metallicity and abundance gradients, with little power to 
constrain how the gradients are distributed about the mean. 
However, by being able to work with single-spectrum galaxies, 
we potentially can examine a much larger sample of galaxies than
previously possible.

Relative to previous studies, our measured gradients in [Fe/H] are
similar but on the whole slightly shallower, while our [Mg/Fe]
gradient matches all the compared studies by being flat. In terms of
stellar mass dependence, we see a flattening of metallicity gradients
as mass increases in line with the compared studies. We find fairly
similar central values for [$\alpha$/Fe] (although slightly smaller),
and also observe an increase in central [$\alpha$/Fe] with stellar
mass as the other studies do. Our metallicity also matches previous
studies both in values and trends with mass.

In sum, we find that using this new technique to find the metallicity
inside an annulus of averaged galaxies roughly agrees with both
observations of individual galaxies and simulated predictions of
galaxies that have formed from mergers and/or monolithic
collapse, a conclusion which is supported by our analysis
of mock data to ensure this method is valid.

This technique may be of further use. With this same data set, one
could extend the analysis to include the broader wavelength range
accessible to the SDSS spectrograph than to most integral field
observations. In addition, this technique could be used for some
higher redshift surveys to measure a similar mean gradient for
galaxies at higher redshift, for example in the AGN and Galaxy
Evolution Survey (AGES), GAMA, or the planned DESI Bright Galaxy Survey.

We thank the referee for many helpful comments, especially in directing
us to consider simulations to confirm our method's validity.

Funding for the SDSS and SDSS-II has been provided by the Alfred P. Sloan 
Foundation, the Participating Institutions, the National Science Foundation, 
the U.S. Department of Energy, the National Aeronautics and Space 
Administration, the Japanese Monbukagakusho, the Max Planck Society, and the 
Higher Education Funding Council for England. The SDSS Web Site is 
http://www.sdss.org/.

The SDSS is managed by the Astrophysical Research Consortium for the 
Participating Institutions. The Participating Institutions are the American 
Museum of Natural History, Astrophysical Institute Potsdam, University of 
Basel, University of Cambridge, Case Western Reserve University, 
University of Chicago, Drexel University, Fermilab, the Institute for 
Advanced Study, the Japan Participation Group, Johns Hopkins University, 
the Joint Institute for Nuclear Astrophysics, the Kavli Institute for 
Particle Astrophysics and Cosmology, the Korean Scientist Group, the 
Chinese Academy of Sciences (LAMOST), Los Alamos National Laboratory, 
the Max-Planck-Institute for Astronomy (MPIA), the Max-Planck-Institute 
for Astrophysics (MPA), New Mexico State University, Ohio State University, 
University of Pittsburgh, University of Portsmouth, Princeton University, 
the United States Naval Observatory, and the University of Washington.

\begin{table*}[t!]
  \centering
  \caption{Gradients in Metallicity Indicators Versus Physical Radius}
  \begin{tabular}{ c | c | c | c | c}
Measure & $\log{M}$ Range & Gradient & Uncertainty in Gradient & Value at 2.5 kpc \\
\hline
$[MgFe]'$ & $10.0 < \log(M) < 10.3$ &  -0.3229 &  0.0165 &  2.4379 \\
$[MgFe]'$ & $10.3 < \log(M) < 10.7$ &  -0.2035 &  0.0071 &  2.8636 \\
$[MgFe]'$ & $10.7 < \log(M) < 11.0$ &  -0.1526 &  0.0050 &  3.1023 \\
$[MgFe]'$ & $11.0 < \log(M) < 11.5$ &  -0.1385 &  0.0059 &  3.2685 \\
$<Fe>$ & $10.0 < \log(M) < 10.3$ &  -0.2493 &  0.0274 &  2.3472 \\
$<Fe>$ & $10.3 < \log(M) < 10.7$ &  -0.1262 &  0.0119 &  2.7090 \\
$<Fe>$ & $10.7 < \log(M) < 11.0$ &  -0.0974 &  0.0081 &  2.8277 \\
$<Fe>$ & $11.0 < \log(M) < 11.5$ &  -0.1032 &  0.0093 &  2.8801 \\
Mg b & $10.0 < \log(M) < 10.3$ &  -0.4068 &  0.0182 &  2.4345 \\
Mg b & $10.3 < \log(M) < 10.7$ &  -0.2961 &  0.0081 &  2.9186 \\
Mg b & $10.7 < \log(M) < 11.0$ &  -0.2078 &  0.0059 &  3.3138 \\
Mg b & $11.0 < \log(M) < 11.5$ &  -0.1658 &  0.0071 &  3.6103 \\
C 4668 & $10.0 < \log(M) < 10.3$ &  -1.1205 &  0.0282 &  3.7319 \\
C 4668 & $10.3 < \log(M) < 10.7$ &  -0.7542 &  0.0123 &  5.1096 \\
C 4668 & $10.7 < \log(M) < 11.0$ &  -0.4804 &  0.0092 &  6.0734 \\
C 4668 & $11.0 < \log(M) < 11.5$ &  -0.3499 &  0.0133 &  6.6346 \\
Ca 4227 & $10.0 < \log(M) < 10.3$ &  -0.1652 &  0.0247 &  1.3840 \\
Ca 4227 & $10.3 < \log(M) < 10.7$ &  -0.1241 &  0.0104 &  1.5666 \\
Ca 4227 & $10.7 < \log(M) < 11.0$ &  -0.0910 &  0.0077 &  1.6885 \\
Ca 4227 & $11.0 < \log(M) < 11.5$ &  -0.0806 &  0.0100 &  1.7647 \\
Corrected H$\beta$ & $10.0 < \log(M) < 10.3$ &  0.1185 &  0.0210 &  2.1687 \\
Corrected H$\beta$ & $10.3 < \log(M) < 10.7$ &  0.1049 &  0.0095 &  2.0620 \\
Corrected H$\beta$ & $10.7 < \log(M) < 11.0$ &  0.0590 &  0.0084 &  1.8736 \\
Corrected H$\beta$ & $11.0 < \log(M) < 11.5$ &  0.1025 &  0.0099 &  1.8275 \\
\hline
$[Fe/H]$ & $10.0 < \log(M) < 10.3$ &  -0.1171 &  0.0135 &  -0.2301 \\
$[Fe/H]$ & $10.3 < \log(M) < 10.7$ &  -0.0301 &  0.0058 &  -0.0079 \\
$[Fe/H]$ & $10.7 < \log(M) < 11.0$ &  -0.0114 &  0.0039 &  0.0264 \\
$[Fe/H]$ & $11.0 < \log(M) < 11.5$ &  -0.0166 &  0.0043 &  0.0092 \\
$[Mg/Fe]$ & $10.0 < \log(M) < 10.3$ &  0.0407 &  0.0309 &  0.0061 \\
$[Mg/Fe]$ & $10.3 < \log(M) < 10.7$ &  0.0046 &  0.0179 &  -0.0430 \\
$[Mg/Fe]$ & $10.7 < \log(M) < 11.0$ &  -0.0146 &  0.0117 &  -0.0315 \\
$[Mg/Fe]$ & $11.0 < \log(M) < 11.5$ &  0.0165 &  0.0089 &  0.0198 \\
$[C/Fe]$ & $10.0 < \log(M) < 10.3$ &  -0.0602 &  0.0252 &  -0.0543 \\
$[C/Fe]$ & $10.3 < \log(M) < 10.7$ &  -0.0611 &  0.0218 &  -0.0136 \\
$[C/Fe]$ & $10.7 < \log(M) < 11.0$ &  -0.0242 &  0.0136 &  0.0951 \\
$[C/Fe]$ & $11.0 < \log(M) < 11.5$ &  -0.0091 &  0.0108 &  0.1771 \\
\end{tabular}
  \tablecomments{The calculated gradients for each measured metallicity 
indicator and the EZ\_Ages results binned by stellar mass. The units are 
dex $\mathrm{kpc^{-1}}$. }
  \label{table:gradients}
\end{table*}

\begin{table*}[t!]
  \centering
  \caption{Gradients in Metallicity Indicators Versus Scaled Radius}
  \begin{tabular}{ c | c | c | c | c}
Measure & $\log{M}$ Range & Gradient & Uncertainty in Gradient & Value at 0.5 $R_{e}$ \\
\hline
$[MgFe]'$ & $10.0 < \log(M) < 10.3$ &  -0.5658 &  0.0281 &  2.9623 \\
$[MgFe]'$ & $10.3 < \log(M) < 10.7$ &  -0.5865 &  0.0187 &  3.0790 \\
$[MgFe]'$ & $10.7 < \log(M) < 11.0$ &  -0.6961 &  0.0202 &  3.1358 \\
$[MgFe]'$ & $11.0 < \log(M) < 11.5$ &  -1.0237 &  0.0394 &  3.1031 \\
$<Fe>$ & $10.0 < \log(M) < 10.3$ &  -0.4494 &  0.0466 &  2.7456 \\
$<Fe>$ & $10.3 < \log(M) < 10.7$ &  -0.3929 &  0.0312 &  2.8281 \\
$<Fe>$ & $10.7 < \log(M) < 11.0$ &  -0.4828 &  0.0327 &  2.8298 \\
$<Fe>$ & $11.0 < \log(M) < 11.5$ &  -0.7990 &  0.0623 &  2.7386 \\
Mg b & $10.0 < \log(M) < 10.3$ &  -0.7023 &  0.0310 &  3.1003 \\
Mg b & $10.3 < \log(M) < 10.7$ &  -0.8199 &  0.0213 &  3.2489 \\
Mg b & $10.7 < \log(M) < 11.0$ &  -0.9070 &  0.0240 &  3.3797 \\
Mg b & $11.0 < \log(M) < 11.5$ &  -1.1906 &  0.0475 &  3.4294 \\
C 4668 & $10.0 < \log(M) < 10.3$ &  -1.9689 &  0.0480 &  5.5489 \\
C 4668 & $10.3 < \log(M) < 10.7$ &  -2.1291 &  0.0322 &  5.9306 \\
C 4668 & $10.7 < \log(M) < 11.0$ &  -2.1628 &  0.0374 &  6.1929 \\
C 4668 & $11.0 < \log(M) < 11.5$ &  -2.6084 &  0.0891 &  6.2052 \\
Ca 4227 & $10.0 < \log(M) < 10.3$ &  -0.2767 &  0.0420 &  1.6587 \\
Ca 4227 & $10.3 < \log(M) < 10.7$ &  -0.3531 &  0.0274 &  1.7002 \\
Ca 4227 & $10.7 < \log(M) < 11.0$ &  -0.4164 &  0.0314 &  1.7078 \\
Ca 4227 & $11.0 < \log(M) < 11.5$ &  -0.6056 &  0.0671 &  1.6634 \\
Corrected H$\beta$ & $10.0 < \log(M) < 10.3$ &  0.2036 &  0.0358 &  1.9744 \\
Corrected H$\beta$ & $10.3 < \log(M) < 10.7$ &  0.2800 &  0.0249 &  1.9396 \\
Corrected H$\beta$ & $10.7 < \log(M) < 11.0$ &  0.2458 &  0.0342 &  1.8490 \\
Corrected H$\beta$ & $11.0 < \log(M) < 11.5$ &  0.6925 &  0.0663 &  1.9176 \\
\hline
$[Fe/H]$ & $10.0 < \log(M) < 10.3$ &  -0.2127 &  0.0228 &  -0.0437 \\
$[Fe/H]$ & $10.3 < \log(M) < 10.7$ &  -0.0808 &  0.0132 &  0.0176 \\
$[Fe/H]$ & $10.7 < \log(M) < 11.0$ &  -0.1652 &  0.0154 &  -0.0189 \\
$[Fe/H]$ & $11.0 < \log(M) < 11.5$ &  -0.1966 &  0.0294 &  -0.0446 \\
$[Mg/Fe]$ & $10.0 < \log(M) < 10.3$ &  0.1090 &  0.0483 &  -0.0537 \\
$[Mg/Fe]$ & $10.3 < \log(M) < 10.7$ &  -0.0067 &  0.0369 &  -0.0473 \\
$[Mg/Fe]$ & $10.7 < \log(M) < 11.0$ &  -0.0689 &  0.0558 &  -0.0326 \\
$[Mg/Fe]$ & $11.0 < \log(M) < 11.5$ &  0.0902 &  0.0653 &  0.0295 \\
$[C/Fe]$ & $10.0 < \log(M) < 10.3$ &  -0.1611 &  0.0469 &  0.0352 \\
$[C/Fe]$ & $10.3 < \log(M) < 10.7$ &  -0.1965 &  0.0410 &  0.0648 \\
$[C/Fe]$ & $10.7 < \log(M) < 11.0$ &  -0.0777 &  0.0601 &  0.1057 \\
$[C/Fe]$ & $11.0 < \log(M) < 11.5$ &  -0.3059 &  0.0717 &  0.1476 \\
\end{tabular}
  \tablecomments{The calculated gradients for each measured metallicity 
indicator and the EZ\_Ages results binned by stellar mass. The units are dex.}
  \label{table:gradients_re}
\end{table*}

\begin{table*}[t!]
  \centering
  \caption{Gradients in Metallicity Indicators Versus Physical Radius}
  \begin{tabular}{ c | c | c | c | c}
Measure & $\sigma$ Range (km s$^{-1}$) & Gradient & Uncertainty in Gradient & Value at 2.5 kpc \\
\hline
$[MgFe]'$ & $30 < \sigma < 125$ &  -0.1152 &  0.0204 &  2.5996 \\
$[MgFe]'$ & $125 < \sigma < 185$ &  -0.1336 &  0.0082 &  2.8832 \\
$[MgFe]'$ & $185 < \sigma < 230$ &  -0.1307 &  0.0066 &  3.0963 \\
$[MgFe]'$ & $230 < \sigma < 325$ &  -0.1185 &  0.0069 &  3.3224 \\
$<Fe>$ & $30 < \sigma < 125$ &  -0.1281 &  0.0341 &  2.4848 \\
$<Fe>$ & $125 < \sigma < 185$ &  -0.1060 &  0.0137 &  2.7160 \\
$<Fe>$ & $185 < \sigma < 230$ &  -0.1042 &  0.0108 &  2.8100 \\
$<Fe>$ & $230 < \sigma < 325$ &  -0.0833 &  0.0109 &  2.9161 \\
Mg b & $30 < \sigma < 125$ &  -0.0934 &  0.0230 &  2.6544 \\
Mg b & $125 < \sigma < 185$ &  -0.1679 &  0.0094 &  2.9668 \\
Mg b & $185 < \sigma < 230$ &  -0.1547 &  0.0079 &  3.3209 \\
Mg b & $230 < \sigma < 325$ &  -0.1443 &  0.0083 &  3.6861 \\
C 4668 & $30 < \sigma < 125$ &  -0.4376 &  0.0354 &  4.4164 \\
C 4668 & $125 < \sigma < 185$ &  -0.5219 &  0.0142 &  5.2554 \\
C 4668 & $185 < \sigma < 230$ &  -0.3636 &  0.0124 &  6.1250 \\
C 4668 & $230 < \sigma < 325$ &  -0.3249 &  0.0153 &  6.7582 \\
Ca 4227 & $30 < \sigma < 125$ &  -0.0772 &  0.0315 &  1.4471 \\
Ca 4227 & $125 < \sigma < 185$ &  -0.0834 &  0.0121 &  1.5851 \\
Ca 4227 & $185 < \sigma < 230$ &  -0.0760 &  0.0104 &  1.6896 \\
Ca 4227 & $230 < \sigma < 325$ &  -0.0658 &  0.0116 &  1.7962 \\
Corrected H$\beta$ & $30 < \sigma < 125$ &  -0.0255 &  0.0260 &  2.1447 \\
Corrected H$\beta$ & $125 < \sigma < 185$ &  0.0525 &  0.0109 &  2.0780 \\
Corrected H$\beta$ & $185 < \sigma < 230$ &  0.0730 &  0.0102 &  1.9444 \\
Corrected H$\beta$ & $230 < \sigma < 325$ &  0.0908 &  0.0115 &  1.7897 \\
\hline
$[Fe/H]$ & $30 < \sigma < 125$ &  -0.0791 &  0.0163 &  -0.1343 \\
$[Fe/H]$ & $125 < \sigma < 185$ &  -0.0545 &  0.0066 &  -0.0342 \\
$[Fe/H]$ & $185 < \sigma < 230$ &  -0.0193 &  0.0044 &  0.0252 \\
$[Fe/H]$ & $230 < \sigma < 325$ &  -0.0069 &  0.0042 &  0.0232 \\
$[Mg/Fe]$ & $30 < \sigma < 125$ &  0.0694 &  0.0268 &  0.0359 \\
$[Mg/Fe]$ & $125 < \sigma < 185$ &  -0.0051 &  0.0124 &  -0.0248 \\
$[Mg/Fe]$ & $185 < \sigma < 230$ &  0.0071 &  0.0095 &  -0.0125 \\
$[Mg/Fe]$ & $230 < \sigma < 325$ &  0.0154 &  0.0103 &  0.0143 \\
$[C/Fe]$ & $30 < \sigma < 125$ &  -0.0107 &  0.0266 &  -0.0537 \\
$[C/Fe]$ & $125 < \sigma < 185$ &  -0.0606 &  0.0156 &  -0.0056 \\
$[C/Fe]$ & $185 < \sigma < 230$ &  0.0056 &  0.0109 &  0.1232 \\
$[C/Fe]$ & $230 < \sigma < 325$ &  -0.0153 &  0.0079 &  0.2063 \\
\end{tabular}
  \tablecomments{The calculated gradients for each measured metallicity 
indicator and the EZ\_Ages results for the data binned by velocity dispersion. The units 
are dex $\mathrm{kpc^{-1}}$. }
  \label{table:gradients_sigma}
\end{table*}

\begin{table*}[t!]
  \centering
  \caption{Gradients in Metallicity Indicators Versus Scaled Radius}
  \begin{tabular}{ c | c | c | c | c}
Measure & $\sigma$ Range (km s$^{-1}$)& Gradient & Uncertainty in Gradient & Value at 0.5 $R_{e}$ \\
\hline
$[MgFe]$ & $30 < \sigma < 125$ &  -0.2322 &  0.0377 &  2.7716 \\
$[MgFe]$ & $125 < \sigma < 185$ &  -0.4138 &  0.0221 &  3.0104 \\
$[MgFe]$ & $185 < \sigma < 230$ &  -0.6042 &  0.0269 &  3.1210 \\
$[MgFe]$ & $230 < \sigma < 325$ &  -0.7705 &  0.0399 &  3.2333 \\
$<Fe>$ & $30 < \sigma < 125$ &  -0.2667 &  0.0629 &  2.6716 \\
$<Fe>$ & $125 < \sigma < 185$ &  -0.3492 &  0.0370 &  2.8064 \\
$<Fe>$ & $185 < \sigma < 230$ &  -0.5079 &  0.0435 &  2.8166 \\
$<Fe>$ & $230 < \sigma < 325$ &  -0.5760 &  0.0630 &  2.8363 \\
Mg b & $30 < \sigma < 125$ &  -0.1845 &  0.0425 &  2.7957 \\
Mg b & $125 < \sigma < 185$ &  -0.4954 &  0.0253 &  3.1388 \\
Mg b & $185 < \sigma < 230$ &  -0.6881 &  0.0319 &  3.3636 \\
Mg b & $230 < \sigma < 325$ &  -0.9081 &  0.0480 &  3.5929 \\
C 4668 & $30 < \sigma < 125$ &  -0.8821 &  0.0654 &  5.0693 \\
C 4668 & $125 < \sigma < 185$ &  -1.5585 &  0.0384 &  5.7810 \\
C 4668 & $185 < \sigma < 230$ &  -1.6801 &  0.0500 &  6.1940 \\
C 4668 & $230 < \sigma < 325$ &  -2.1148 &  0.0888 &  6.5130 \\
Ca 4227 & $30 < \sigma < 125$ &  -0.1371 &  0.0582 &  1.5715 \\
Ca 4227 & $125 < \sigma < 185$ &  -0.2527 &  0.0327 &  1.6673 \\
Ca 4227 & $185 < \sigma < 230$ &  -0.3539 &  0.0421 &  1.7027 \\
Ca 4227 & $230 < \sigma < 325$ &  -0.4386 &  0.0675 &  1.7414 \\
Corrected H$\beta$ & $30 < \sigma < 125$ &  -0.0446 &  0.0481 &  2.1860 \\
Corrected H$\beta$ & $125 < \sigma < 185$ &  0.1457 &  0.0294 &  2.0196 \\
Corrected H$\beta$ & $185 < \sigma < 230$ &  0.3013 &  0.0414 &  1.9126 \\
Corrected H$\beta$ & $230 < \sigma < 325$ &  0.5337 &  0.0667 &  1.8296 \\
\hline
$[Fe/H]$ & $30 < \sigma < 125$ &  -0.2311 &  0.0303 &  -0.0513 \\
$[Fe/H]$ & $125 < \sigma < 185$ &  -0.0709 &  0.0151 &  0.0185 \\
$[Fe/H]$ & $185 < \sigma < 230$ &  -0.1157 &  0.0176 &  -0.0040 \\
$[Fe/H]$ & $230 < \sigma < 325$ &  -0.1740 &  0.0290 &  -0.0325 \\
$[Mg/Fe]$ & $30 < \sigma < 125$ &  0.1363 &  0.0551 &  -0.0522 \\
$[Mg/Fe]$ & $125 < \sigma < 185$ &  -0.0068 &  0.0377 &  -0.0474 \\
$[Mg/Fe]$ & $185 < \sigma < 230$ &  0.1155 &  0.0411 &  -0.0038 \\
$[Mg/Fe]$ & $230 < \sigma < 325$ &  0.0784 &  0.0567 &  0.0236 \\
$[C/Fe]$ & $30 < \sigma < 125$ &  -0.1861 &  0.0517 &  0.0284 \\
$[C/Fe]$ & $125 < \sigma < 185$ &  -0.2005 &  0.0419 &  0.0616 \\
$[C/Fe]$ & $185 < \sigma < 230$ &  -0.0758 &  0.0575 &  0.1060 \\
$[C/Fe]$ & $230 < \sigma < 325$ &  -0.1251 &  0.0710 &  0.1785 \\
\end{tabular}
  \tablecomments{The calculated gradients for each measured metallicity 
indicator and the EZ\_Ages results binned by velocity dispersion. The units are dex.}
  \label{table:gradients_re_sigma}
\end{table*}

\begin{table}[t!]
  \caption{Logarithmic Gradients: Stellar Mass Binning}
  \begin{tabular}{ c | c | c | c }
Measure & $\log{M}$ & ${\rm d}({\rm Measure})/{\rm d}\log{r}$ & Error \\
\hline
$[Fe/H]$ & $10.0 < \log(M) < 10.3$ &  -0.3028 &  0.0657 \\
$[Fe/H]$ & $10.3 < \log(M) < 10.7$ &  -0.0906 &  0.0296 \\
$[Fe/H]$ & $10.7 < \log(M) < 11.0$ &  -0.1709 &  0.0343 \\
$[Fe/H]$ & $11.0 < \log(M) < 11.5$ &  -0.1832 &  0.0702 \\
$[Mg/Fe]$ & $10.0 < \log(M) < 10.3$ &  0.1552 &  0.2066 \\
$[Mg/Fe]$ & $10.3 < \log(M) < 10.7$ &  -0.0075 &  0.1302 \\
$[Mg/Fe]$ & $10.7 < \log(M) < 11.0$ &  -0.0712 &  0.1519 \\
$[Mg/Fe]$ & $11.0 < \log(M) < 11.5$ &  0.0840 &  0.2212 \\
$[C/Fe]$ & $10.0 < \log(M) < 10.3$ &  -0.2294 &  0.1886 \\
$[C/Fe]$ & $10.3 < \log(M) < 10.7$ &  -0.2204 &  0.1292 \\
$[C/Fe]$ & $10.7 < \log(M) < 11.0$ &  -0.0803 &  0.1836 \\
$[C/Fe]$ & $11.0 < \log(M) < 11.5$ &  -0.2851 &  0.2300 \\
\end{tabular}
  \tablecomments{The calculated logarithmic gradients for the metallicity and 
[$\alpha$/Fe] ratios for all four stellar mass bins.}
  \label{table:loggrads}
\end{table}

\begin{table}[t!]
  \caption{Logarithmic Gradients: Velocity Dispersion Binning}
  \begin{tabular}{ c | c | c | c }
Measure & $\sigma$ & $d({\rm Measure})/d\log{r}$ & Error \\
\hline
$[Fe/H]$ & $30 < \sigma < 125$ &  -0.3069 &  0.0853 \\
$[Fe/H]$ & $125 < \sigma < 185$ &  -0.0708 &  0.0343 \\
$[Fe/H]$ & $185 < \sigma < 230$ &  -0.1203 &  0.0403 \\
$[Fe/H]$ & $230 < \sigma < 325$ &  -0.1866 &  0.0776 \\
$[Mg/Fe]$ & $30 < \sigma < 125$ &  0.1810 &  0.2223 \\
$[Mg/Fe]$ & $125 < \sigma < 185$ &  -0.0068 &  0.1312 \\
$[Mg/Fe]$ & $185 < \sigma < 230$ &  0.1201 &  0.1250 \\
$[Mg/Fe]$ & $230 < \sigma < 325$ &  0.0840 &  0.2212 \\
$[C/Fe]$ & $30 < \sigma < 125$ &  -0.2472 &  0.1957 \\
$[C/Fe]$ & $125 < \sigma < 185$ &  -0.2005 &  0.1293 \\
$[C/Fe]$ & $185 < \sigma < 230$ &  -0.0788 &  0.1791 \\
$[C/Fe]$ & $230 < \sigma < 325$ &  -0.1341 &  0.2528 \\
\end{tabular}
  \tablecomments{The calculated logarithmic gradients for the metallicity and 
[$\alpha$/Fe] ratios for all four velocity dispersion bins.}
  \label{table:loggrads_sigma}
\end{table}

\clearpage

\bibliographystyle{apj} 
\bibliography{Metallicity_Gradient}

\end{document}